\documentclass[lettersize,journal]{IEEEtran}
\usepackage{amsmath,amsfonts}
\usepackage{algorithmic}
\usepackage{algorithm}
\usepackage{array}
\usepackage[caption=false,font=normalsize,labelfont=sf,textfont=sf]{subfig}
\usepackage{textcomp}
\usepackage{stfloats}
\usepackage{url}
\usepackage{verbatim}
\usepackage{graphicx}
\usepackage{cite}
\usepackage{xcolor}
\usepackage{hyperref}
\usepackage{cleveref}
\hypersetup{colorlinks,linkcolor={black},citecolor={blue},urlcolor={blue}} 
\def\BibTeX{{\rm B\kern-.05em{\sc i\kern-.025em b}\kern-.08em
    T\kern-.1667em\lower.7ex\hbox{E}\kern-.125emX}}
\usepackage{balance}
\begin{document}
\title{Nonlinear Feedback Linearization and LQG/LTR Control: A Comparative Study for a Single-Machine Infinite-Bus System}
\author{Pratik Vernekar
\thanks{P. Vernekar is with General Motors, Milford, MI 48380, USA (e-mail: pratik.vernekar@gmail.com).}}
\markboth{Nonlinear Feedback Linearization and LQG/LTR Control: A Comparative Study for a Single-Machine Infinite-Bus System}
{}

\maketitle

\begin{abstract}
This paper presents a comparative study of three advanced control strategies for a single-machine infinite-bus (SMIB) system: the nonlinear feedback linearizing controller (NFLC), the integral-NFLC (INFLC), and the linear-quadratic-Gaussian/loop transfer recovery (LQG/LTR) control. The NFLC and INFLC techniques use exact feedback linearization to precisely cancel the SMIB system nonlinearities, enabling the use of decentralized, linear, and optimal controllers for the decoupled generator and turbine-governor systems while remaining unaffected by the SMIB system's internal dynamics and operating conditions. In contrast, the LQG/LTR approach employs an enhanced Kalman filter, designed using the LTR procedure and a detailed frequency-domain loop-shaping analysis, to achieve a reasonable trade-off between optimal performance, noise/disturbance rejection, robustness recovery, and stability margins for the SMIB system. We provide a control synthesis framework for constructing practical, verifiable, scalable, and resilient linear and nonlinear controllers for SMIB and multi-machine power systems by utilizing a high-fidelity plant model for validation, a reduced-order control-design model, and the correlations between the two models' control inputs. Rigorous simulations and comparative analysis of the proposed controllers and a full-state linear–quadratic regulator show the benefits, constraints, and trade-offs of each controller in terms of transient response, steady-state error, robustness, rotor angle stability, frequency control, and voltage regulation under different operating conditions. Ultimately, this study aims to guide the selection of appropriate control strategies for large-scale power systems, enhancing the overall resilience and reliability of the electric grid.
\end{abstract}


\section{Introduction}
\IEEEPARstart{T}{he} modern power system, also known as the smart grid, is a highly complex, widely interconnected, large-scale nonlinear system. The stable and reliable operation of power systems has become increasingly challenging with the rapid integration of intermittent renewable energy sources like solar panels and wind turbines, distributed generation, energy storage systems, complex load profiles like plug-in electric vehicles, and advanced communication technologies into the smart grid. The overall stability of the power grid can be decomposed into the following objectives: a) automatic voltage regulation via excitation control; b) rotor angle stability and load frequency control (LFC) via turbine-governor control; and c) transient stability enhancement via coordinated control in the event of severe faults, external disturbances, and system overload. In addition to multi-machine benchmark models like the IEEE 10-machine 39-bus New England test system \cite{ref1}, the single-machine infinite-bus (SMIB) system is often used to develop robust control strategies and conduct stability studies on the power grid because it captures the essential dynamics of the generator, turbine-governor system, and power network.

Many classical and advanced control strategies have been developed so far for the effective control and stability enhancement of SMIB and multi-machine power systems. Power system stabilizers have been widely used to enhance the power system's transient stability in response to minor disturbances by adding damping via excitation control modulation \cite{ref2}. Novel control approaches using sliding mode control (SMC) \cite{ref3,ref4,ref5,ref6} are used for the LFC design in power systems. The control methods in \cite{ref2,ref3,ref4,ref5,ref6} are based on reduced-order power system models linearized around specific operating points. Therefore, these techniques cannot provide adequate damping and preserve transient stability over a wide range of operating conditions, particularly when power systems experience severe disturbances like three-phase short-circuit faults or sudden fluctuations in load demands. Also, the control approaches in \cite{ref2,ref3,ref4,ref5,ref6} are not validated on high-fidelity plant models with uncertainties, which are closer reflections of practical multi-machine power systems. Furthermore, the control input chattering, saturation, and singularity problems in SMC-based LFC approaches \cite{ref3,ref4,ref5,ref6} make them unfeasible for implementation on higher-order nonlinear power system models.

Nonlinear controllers, conversely, are unaffected by operating conditions and significantly enhance power system stability and transient performance across a broad spectrum of operating points, even amid severe disturbances or faults \cite{ref7,ref8,ref9,ref10,ref11,ref12,ref13,ref14,ref15,ref16,ref17,ref18,ref19,ref20,ref21,ref22}. The feedback linearization technique, which can be broadly classified into three distinct forms, namely, direct feedback linearization (DFL), partial feedback linearization (PFL), and exact feedback linearization (EFL), has been widely used to design excitation controllers for SMIB and multi-machine power systems \cite{ref7,ref8,ref9,ref10,ref11,ref12,ref13,ref14,ref15,ref16,ref17,ref18}. The DFL methods proposed in \cite{ref7,ref8,ref9,ref10} necessitate the measurement of rotor angle, real and reactive power, real-time calculation of higher-order derivatives, and virtual control input design to cancel the system nonlinearities, which reduces the controller's robustness in the presence of measurement noise and external perturbations. Several PFL techniques for SMIB and multi-machine power systems have been presented to address the DFL limitations \cite{ref13,ref14,ref15,ref16,ref17,ref18}. In \cite{ref13,ref14,ref15,ref16}, the synchronous generator's speed is used for the PFL design, avoiding the challenges of rotor angle measurement and differentiator design in the DFL \cite{ref7,ref8,ref9,ref10} and PFL procedures \cite{ref17,ref18}. The PFL approach only partially linearizes higher-order dynamics in detailed generator models with many coupled differential equations. The remaining nonlinear dynamics, also known as the zero or internal dynamics, are not asymptotically stable in higher-fidelity power system models. Thus, employing the PFL controller to stabilize the zero dynamics of SMIB and multi-machine power systems with multiple interconnected generators is challenging due to the increased complexity and interdependencies. The PFL control schemes, which rely heavily on an accurate power system model, fail to achieve the desired stability or transient response in the presence of parametric variations in inertia, damping, transmission line reactance, and transient characteristics under different operating conditions and lack the robustness required to handle variations, severe faults, and uncertainties effectively. Adaptive and backstepping control techniques, such as robust adaptive backstepping \cite{ref19}, extended backstepping \cite{ref20}, and fractional-order SMC-based backstepping \cite{ref21}, have been proposed to mitigate the drawbacks of DFL and PFL techniques and stabilize SMIB and multi-machine power systems. However, applying advanced adaptive and backstepping control techniques to power systems presents some challenges, such as parameter estimation and identification problems under disturbances and noise leading to incorrect adaptation and degraded control performance, and excitation of unmodeled dynamics or high-frequency modes of the power system resulting in undesired oscillations.

Motivated by the above discussion, this paper presents two nonlinear feedback linearizing control (NFLC) techniques and the linear-quadratic-Gaussian/loop transfer recovery (LQG/LTR) control strategy to enhance the transient response, stability, and robustness of the SMIB system. The NFLC methods effectively address the nonlinear and internal dynamics of the SMIB system, improve transient stability and performance during severe faults or load changes, and provide good post-fault voltage regulation. The LQG/LTR technique achieves optimal control, enhances the system's robustness under model uncertainties and disturbances, enables loop shaping of the transfer function for improved control over the system's frequency response and stability, and estimates the unmeasurable states of the SMIB system using a Kalman filter. This study aims to develop robust linear and nonlinear control strategies for the SMIB system across diverse operating conditions and disturbances that can be adapted to different power system configurations, providing greater control design flexibility. By exploring these advanced techniques, we seek to expand the control design toolkit for power engineers, enhancing grid resilience and efficiency.

Most controllers designed for SMIB and multi-machine power systems are based on the classical third-order one-axis generator model or simplified reduced-order power system models \cite{ref2,ref3,ref4,ref5,ref6,ref7,ref8,ref9,ref10,ref11,ref12,ref13,ref14,ref16,ref20,ref21,ref22,ref23}. Also, the turbine-governor dynamics are not considered in the design of most controllers reported in the literature. Higher-order power system models, including a two-axis fifth-order generator model in \cite{ref15,ref19,ref24}, a sixth-order generator-turbine model in \cite{ref17}, and a ninth-order SMIB system model with a seventh-order generator and second-order hydraulic turbine in \cite{ref18}, are used for controller design in some studies. We use first principles to develop a high-fidelity ninth-order model of the SMIB system, which comprises the synchronous generator, turbine, governor, transient and sub-transient flux linkages, and exciter system dynamics. This high-fidelity model serves as the plant or validation model and is not used for the controller design. We then present a fifth-order control-design model (CDM) of the SMIB system, which comprises the third-order one-axis generator model and a second-order turbine-governor model. The reduced-order CDM captures the main features of the high-fidelity model and is simple enough for robust controller design and stability studies of the SMIB system.

First, the EFL technique \cite{ref25} is used to design two nonlinear controllers: a nonlinear feedback linearizing controller (NFLC) and an integral-NFLC (INFLC). This method uses nonlinear coordinate transformations to convert the original nonlinear power system model into two reduced-order, decoupled, fully linear subsystems by precisely canceling nonlinearities and selecting appropriate, measurable SMIB states as system outputs. Linear decentralized state feedback controllers are designed separately for the two decoupled linear subsystems using optimal control theory. Also, the INFLC method uses integral action to reduce the steady-state error. The EFL-based solution minimizes the challenge of stabilizing the unstable internal dynamics of SMIB and interconnected multi-machine power systems, unlike PFL-based approaches \cite{ref13,ref14,ref15,ref16,ref17,ref18}. The NFLC and INFLC techniques outperform linear controllers \cite{ref2,ref3,ref4,ref5,ref6}, regardless of operating conditions, by simultaneously controlling the rotor angle, speed, and terminal voltage of the SMIB system while enhancing overall system stability and damping under various real-world scenarios. Applying the proposed nonlinear control strategies to the high-fidelity SMIB system model necessitates estimating the rotor flux linkages and non-physical $q$-axis voltage of the reduced-order CDM. We provide a mechanism to reconstruct the non-physical, unmeasurable states of the reduced-order CDM from the measurable states of the high-fidelity model, resulting in an interconnection between the control inputs of the two SMIB system models of differing complexity. However, the NFLC and INFLC methods have certain practical constraints, including the need for precise knowledge of system nonlinearities, rotor angle measurement, and sensitivity to plant and model mismatch, uncertainties, and measurement noise.

Next, to improve the robustness and mitigate the drawbacks of the proposed NFLC and INFLC strategies, we propose a robust LQG/LTR control technique \cite{ref26}, which has been used to reduce inter-area oscillations in multi-machine power systems \cite{ref23,ref24}. The LQG/LTR method utilizes an enhanced Kalman filter to estimate the rotor angle, the $q$-axis voltage, and the remaining unmeasurable states of the SMIB system. Because real-time measurement of the rotor angle of the synchronous generator is difficult in an interconnected multi-machine power system, we use the synchronous generator's speed, which can be measured directly using speed sensors, along with the terminal generator voltage, as system outputs for the LQG/LTR controller design. Also, the speed, which is directly related to the derivative of the rotor angle, will provide more damping to the system when used as output feedback like the PFL-based approaches \cite{ref13,ref14,ref15,ref16}. The LQG/LTR technique employs covariance matrices reflecting fictitious process and measurement noise intensities as Kalman filter design parameters to account for uncertainties, unmodeled dynamics, and external disturbances. The Kalman filter gains are appropriately tuned using the LTR procedure to recover the robustness features of the full-state feedback linear quadratic regulator (LQR) at the plant input and achieve a reasonable trade-off between noise/disturbance rejection, closed-loop stability margin, and nominal system performance. The LQG controller gains are designed using optimal control theory to minimize the LQG cost function and maintain the SMIB system's closed-loop stability. A detailed frequency domain loop-shaping analysis of the proposed LQG/LTR controller uses the loop transfer functions of the fifth-order linear CDM of the SMIB system. 

Finally, the NFLC, INFLC, and LQG/LTR controllers are validated on the reduced-order CDM and high-fidelity SMIB system plant models under different operating conditions, including a three-phase short-circuit fault at the generator terminal and variations in system load. We perform a comprehensive comparative analysis of the proposed controllers and a full-state feedback LQR in terms of stability, transient response, steady-state error, and robustness under various practical operating scenarios. Our findings, based on rigorous simulations and comparative studies, provide insights into the strengths, limitations, trade-offs, and practical considerations associated with each control strategy, thereby aiding practitioners in selecting the most suitable approach for SMIB and multi-machine power system control. To the authors' best knowledge, the control synthesis methodology and comparative analysis of the proposed linear and nonlinear controllers in this paper are yet to be presented for the robust and optimal control of SMIB and multi-machine power systems.

The main contributions of this paper are as follows:

1) Using a high-fidelity ninth-order physics-based plant model and a fifth-order CDM of the SMIB system, we provide a framework for developing realistic, feasible, empirical, and robust linear and nonlinear controllers for the SMIB system, which can be adapted to develop control algorithms for multi-machine power systems and complex smart grids.

2) We reconstruct the reduced-order CDM's non-physical, unmeasurable states from the measurable states of the high-fidelity plant model for the real-time application of NFLC and INFLC strategies on the SMIB system. These NFLC and INFLC approaches enable decentralized, linear, and optimal control of the decoupled generator and turbine-governor subsystems, regardless of the internal dynamics and operating conditions of the SMIB system.

3) The proposed LQG/LTR technique gives a systematic framework for robust linear control design for SMIB and multi-machine power systems. A detailed frequency domain loop-shaping design analysis of the LQG/LTR procedure is performed to obtain a reasonable trade-off between noise/disturbance rejection, robustness recovery, and stability margins for the SMIB system. 

4) Extensive simulations compare the NFLC, INFLC, LQG/LTR, and full-state LQR controllers in terms of transient response, convergence rate, steady-state error, and robustness under different real-world operating scenarios such as severe power system faults and load variations.

\section{High-Fidelity Model of the SMIB System}

\begin{figure}[!t]
	\centering
	\includegraphics[height = 4.2cm, width = 8.2cm]{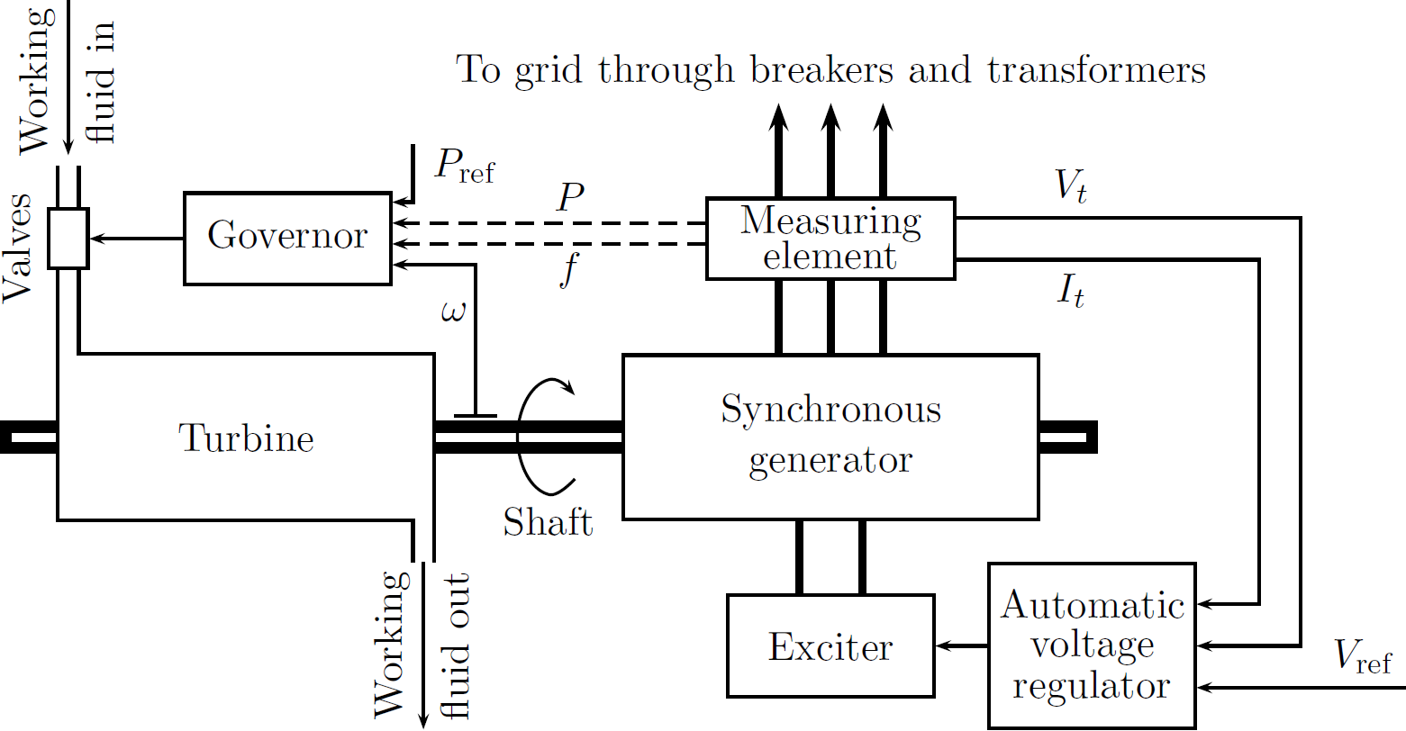}
	\caption{Structure of a single-generation unit.}
	\label{fig1}
\end{figure}

Fig. \ref{fig1} depicts a single-generation unit with a synchronous generator powered by a turbine-governor system \cite{ref27}. In an interconnected power system, where a synchronous generator is connected to a power grid, there are two control loops: the automatic voltage regulator (AVR) loop and the load frequency control (LFC) loop. The controllers are set for a particular operating condition and accommodate changes in the load demand to maintain the frequency and voltage magnitude within the specified limits. Small changes in real power depend on changes in the rotor angle $\delta $, and thus the frequency $\omega $. The reactive power depends on the voltage magnitude, i.e., the generator excitation. The synchronous generator consists of the following components: a) a stator with three-phase armature windings that supply power to the grid, b) a hydraulic turbine-driven rotor with an excitation field winding wrapped around it, and c) two short-circuit damper windings to dampen the rotor's mechanical oscillations. 

\begin{figure}[!t]
	\centering
	\includegraphics[height = 5.7cm, width = 6.5cm]{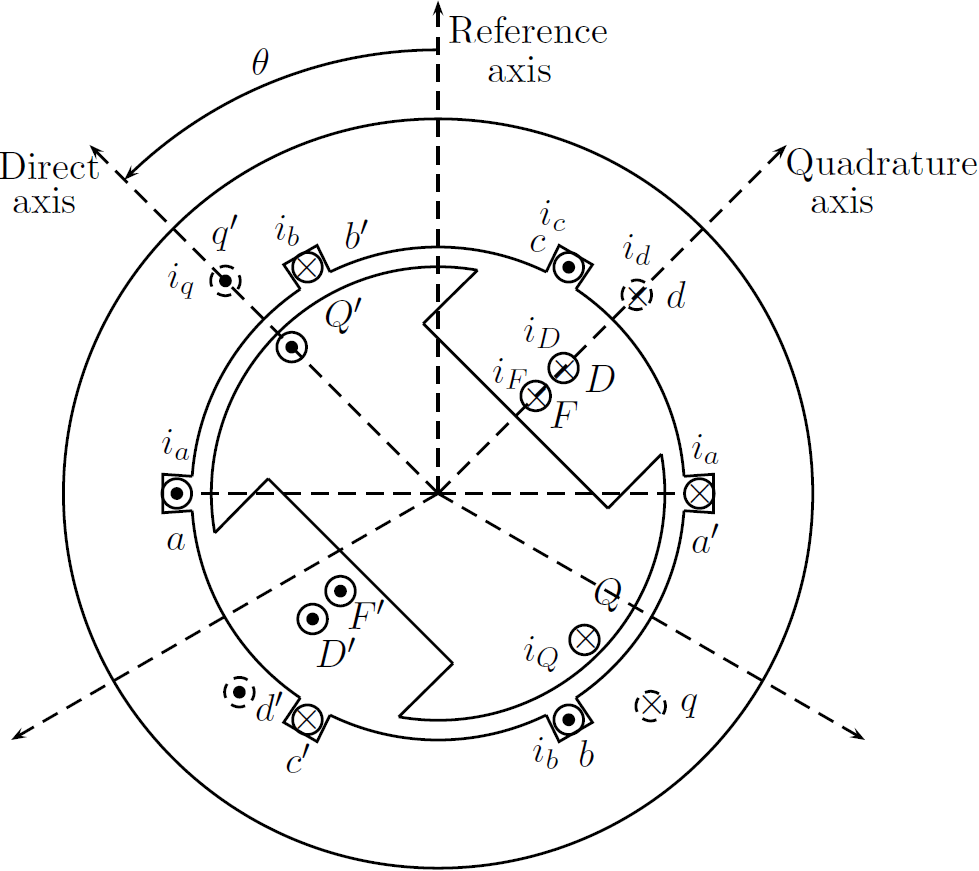}
	\caption{A synchronous generator schematic with the reference directions.}
	\label{fig2}
\end{figure}

A schematic of a synchronous generator with the reference directions is shown in Fig. \ref{fig2} \cite{ref28}. The armature winding, which carries the load current $I_t$ and supplies power to the grid, is placed in equidistant slots on the inner surface of the stator and consists of three identical phase windings, namely, $aa'$, $bb'$, and $cc'$. The direct current (DC) excitation winding represented by $FF'$ is wrapped around the rotor. The two short-circuit damper or amortisseur windings represented by $DD'$ and $QQ'$ help to dampen the rotor's mechanical oscillations. We define $\mathbf{v} = [v_a, v_b, v_c, -v_F, -v_D, -v_Q]^\mathrm{T}$ as the voltage vector consisting of the three-phase terminal voltages ($v_a$, $v_b$, $v_c$), the field winding voltage ($v_F$), and the voltages of the two damper windings ($v_D$, $v_Q$). The corresponding current vector is defined as $\mathbf{i} = [i_a, i_b, i_c, i_F, i_D, i_Q]^\mathrm{T}$.  

A synchronous generator connected to an infinite bus through a transmission line with resistance $R_e$ and inductance $L_e$ is illustrated in \cite{ref2,ref7,ref9,ref16,ref18,ref22}. An infinite bus is an approximation of a large-scale interconnected power system where the action of a single generator will not affect the operation of the power grid. Additionally, a single-machine infinite-bus (SMIB) system, which is frequently employed for stability studies of multi-machine power systems, qualitatively manifests the principal features of a large interconnected power grid composed of numerous synchronous machines.
The electrical dynamics of the synchronous generator are obtained by using Park's transformation to transform variables from the static $abc$ frame to a synchronously rotating $dq$ frame. The infinite bus constraints are as follows \cite{ref28}:
\begin{equation}
	\scalebox{0.9}{$
	\begin{bmatrix} v_d \\ v_q \end{bmatrix} = R_e \begin{bmatrix} i_d \\ i_q \end{bmatrix} + L_e \begin{bmatrix} \dot{i_d} \\ \dot{i_q}
\end{bmatrix} - \omega L_e \begin{bmatrix} -i_q \\ i_d \end{bmatrix} + \sqrt{3}V_{\infty}\begin{bmatrix} -\sin{(\delta - \alpha)} \\
	\cos{(\delta - \alpha)}\end{bmatrix} $},
	\label{inf_bus}
\end{equation}
in which $\alpha$ is the load angle, $V_{\infty}$ is the infinite bus voltage, and $v_d$ and $v_q$ are the voltages along the synchronously rotating direct $d$-axis and quadrature $q$-axis in the $dq$ frame, respectively. Similarly, $i_d$ and $i_q$ are the $d$- and $q$-axis currents in the $dq$ frame, as shown in Fig. \ref{fig2}. Using \eqref{inf_bus}, the electrical dynamics of a synchronous generator connected to an infinite bus can be written as follows:
\begin{equation}
	\scalebox{0.9}{$
	\begin{aligned}	
		&\begin{bmatrix} L_d+L_e & kM_F & kM_D & 0 & 0 \\
		-kM_F & -L_F & -M_R & 0 & 0 \\
		-kM_D & -M_R & -L_D & 0 & 0 \\
		0 & 0 & 0 & L_q+L_e & kM_Q \\
		0 & 0 & 0 & -kM_Q & -L_Q \end{bmatrix} \begin{bmatrix} \dot{i_d} \\ \dot{i_F} \\
		\dot{i_D} \\ \dot{i_q} \\ \dot{i_Q} \end{bmatrix} =\\
		&\begin{bmatrix} -r-R_e & 0 & 0 & -\omega (L_q+L_e) & -\omega kM_Q \\
		0 & r_F & 0 & 0 & 0\\
		0 & 0 & r_D & 0 & 0\\
		\omega (L_d+L_e) & \omega kM_F & \omega kM_D & -r-R_e & 0 \\
		0 & 0 & 0 & 0 & r_Q \end{bmatrix} \begin{bmatrix} i_d \\ i_F \\ i_D \\ i_q \\ i_Q \end{bmatrix} \\ 
		&- \begin{bmatrix} 0 \\ v_F \\ 0 \\ 0 \\ 0 \end{bmatrix} - \sqrt{3}V_\infty\begin{bmatrix} -\sin{(\delta - \alpha)} \\ 0 \\ 0 \\ 
		\cos{(\delta - \alpha)} \\ 0 \end{bmatrix},
	\end{aligned}$}\\
	\label{electrical_dynamics}
\end{equation}
in which $k = \sqrt{3/2}$ is the flux linkage constant, $M_F$ is the mutual inductance between the $dd'$ and $FF'$ windings, $M_D$ is the mutual inductance between the $dd'$ and $DD'$ windings, $M_Q$ is the mutual inductance between the $qq'$ and $QQ'$ windings, $M_R$ is the mutual inductance between the $FF'$ and $DD'$ windings, and $L_d$, $L_F$, $L_D$, $L_q$, $L_Q$ are the self-inductances of the $dd'$, $FF'$, $DD'$, $qq'$, $QQ'$ windings, respectively. Also, $r$ is the resistance of each of the three-phase stator windings ($aa'$, $bb'$, $cc'$), and $r_F$, $r_D$, and $r_Q$ are the resistances of the $FF'$, $DD'$, and $QQ'$ windings, respectively. Note that all the equations used in the derivation of the high-fidelity model are in the per-unit (p.u.) system.

Next, the swing equation gives the mechanical dynamics of the synchronous generator in the p.u. system as follows \cite{ref28}:
\begin{equation}
	\scalebox{0.97}{$
	\begin{aligned}
		\dot{\omega} =& -\frac{(L_d i_d i_q + kM_F i_F i_q +
			kM_Di_Di_q - L_qi_di_q - kM_Qi_di_Q)}{3\tau _j}\\ &- \frac{1}{\tau _j}D\omega +
		\frac{1}{\tau _j}T_m, \ \ \ \
		\dot{\delta} = \omega - 1, \\
	\end{aligned}$}
	\label{mechanical_pu}
\end{equation}
in which $D$ is the damping constant, $T_m$ is the mechanical torque of the turbine, $\tau _j=2H\omega _R$, where $H$ is the inertia constant, and $\omega _R$ is the base/reference angular velocity.

The dynamics of the turbine-governor system are given next \cite{ref7}. The dynamics of the turbine in the p.u. system are
\begin{equation}
	\dot{T}_m=-\frac{1}{\tau _T}T_m+\frac{K_T}{\tau _T}G_V,
	\label{turbine}
\end{equation}              
in which $G_V$ is the gate opening of the turbine, $\tau _T$ is the time constant of the turbine, and $K_T$ is the gain of the turbine. The dynamics of the governor in the p.u. system are
\begin{equation}
	\dot{G}_{V}=-\frac{1}{\tau _G}G_V+\frac{K_G}{\tau _G}\bigg{(}u_T-\frac{\omega }{R_T}\bigg{)},
	\label{governor}
\end{equation}
in which $u_T$ is the turbine valve control, $\tau _G$ is the time constant of the speed governor, $K_G$ is the gain of the speed governor, and $R_T$ is the regulation constant.

By combining the electrical and mechanical dynamics of the synchronous generator given in \eqref{electrical_dynamics} and \eqref{mechanical_pu}, with the turbine-governor dynamics described in \eqref{turbine} and \eqref{governor}, we obtain the ninth-order high-fidelity model of the SMIB system in the nonlinear state-space form as follows:
\begin{equation}
\scalebox{0.97}{$
	\begin{aligned}
		\dot{I}_d= & \ F_{11}I_d + F_{12}I_F+F_{13}I_D+F_{14}I_q\omega+F_{15}I_Q\omega \\ 
		& +F_{16}\sin(\delta -\alpha)+G_{11}V_F,\\
		\dot{I}_F= & \ F_{21}I_d+F_{22}I_F+F_{23}I_D + F_{24}I_q\omega  +F_{25}I_Q\omega \\
		& + F_{26}\sin(\delta -\alpha )+G_{21}V_F,\\
		\dot{I}_D= & \ F_{31}I_d+F_{32}I_F+F_{33}I_D+ F_{34}I_q\omega +F_{35}I_Q\omega \\
		& + F_{36}\sin(\delta -\alpha )+G_{31}V_F,\\
		\dot{I}_q= & \ F_{41}I_d\omega + F_{42}I_F\omega +F_{43}I_D\omega +F_{44}I_q +F_{45}I_Q \\ 
		& + F_{46}\cos(\delta -\alpha ),\\
		\dot{I}_Q= & \ F_{51}I_d\omega  +F_{52}I_F\omega +F_{53}I_D\omega  +F_{54}I_q+F_{55}I_Q \\ 
		& +F_{56}\cos(\delta -\alpha ),\\
		\dot{\omega }=& \ F_{61}I_dI_q + F_{62}I_FI_q + F_{63}I_DI_q +F_{64}I_dI_Q+ F_{65}\omega \\
		& +F_{66}T_m,\\
		\dot{\delta }= & \ \omega - 1,\\
		\dot{T}_m= & \ F_{81}T_m+F_{82}G_V,\\
		\dot{G}_V= & \ F_{91}\omega +F_{92}G_V+G_{92}u_T,\\
	\end{aligned}$}
	\label{ninthorder}
\end{equation}
in which $F_{11}=-L_{d1}(r+R_e)$, $F_{12}=kM_{F1}r_F$, $F_{13}=kM_{D1}r_D$, $F_{14}=-(L_q+L_e)L_{d1}$, $F_{15}=-kM_QL_{d1}$, $F_{16}= V_\infty L_{d1}$, $G_{11}=- kM_{F1}$, $F_{21}=kM_{F1}(r+R_e)$, $F_{22}=- L_{F1}r_F$, $F_{23}=- M_{R1}r_D$, $F_{24}=kM_{F1}(L_q+L_e)$, $F_{25}=k^2M_{F1}M_Q$, $F_{26}=-V_\infty kM_{F1}$, $G_{21}=L_{F1}$, $F_{31}=kM_{D1}(r+R_e)$, $F_{32}=-M_{R1}r_F$, $F_{33}=-L_{D1}r_D$, $F_{34}=kM_{D1}(L_q+L_e)$, $F_{35}=k^2M_{D1}M_Q$, $F_{36}= -V_\infty kM_{D1}$, $G_{31}=M_{R1}$, $F_{41}=L_{q1}(L_d+L_e)$, $F_{42}=kM_FL_{q1}$, $F_{43}=kM_DL_{q1}$, $F_{44}=- L_{q1}(r+R_e)$, $F_{45}= kM_{Q1}r_Q$, $F_{46}=-V_\infty L_{q1}$, $F_{51}=-kM_{Q1}(L_d+L_e)$, $F_{52}=- k^2M_{Q1}M_F$, $F_{53}=-k^2M_{Q1}M_D$, $F_{54}=kM_{Q1}(r+R_e)$, $F_{55}=- L_{Q1}r_Q$, $F_{56}= V_\infty kM_{Q1}$, $F_{61}=-\frac{1}{\tau _j}(L_d-L_q)$, $F_{62}=-\frac{1}{\tau _j}kM_F$, $F_{63}=-\frac{1}{\tau _j}kM_D$, $F_{64}=\frac{1}{\tau _j}kM_Q$, $F_{65}=-\frac{1}{\tau _j}D$, $F_{81}=-\frac{1}{\tau _T}$, $F_{82}=\frac{K_T}{\tau _T}$,
$F_{91}=-\frac{K_G}{\tau _GR_T}$, $F_{92}=-\frac{1}{\tau _G}$, $G_{92}=\frac{K_G}{\tau _G}$, $L_{d1} = \frac{1}{\mu }(M_R^2 - L_DL_F)$, $L_{F1} = \frac{1}{\mu }(M_D^2k^2-L_D(L_d+L_e))$, $L_{D1} = \frac{1}{\mu }(M_F^2k^2-L_F(L_d+L_e))$, $M_{F1} = \frac{1}{\mu }(M_DM_R - L_DM_F)$, $M_{D1} = \frac{1}{\mu }(M_FM_R - L_FM_D)$, $M_{R1} = \frac{1}{\mu }((L_d + L_e)M_R - M_DM_Fk^2)$, $L_{q1} = \frac{L_Q}{\nu }$, $L_{Q1} = \frac{L_e+L_q}{\nu }$, $M_{Q1}=\frac{M_Q}{\nu }$, $\mu = (L_d + L_e)M_R^2  - L_DL_F(L_d + L_e) + k^2(L_DM_F^2 + L_FM_D^2 - 2M_DM_FM_R)$ and $\nu = -k^2M_Q^2 + L_Q(L_e+L_q)$. In \eqref{ninthorder}, the state variables $i_d$, $i_F$, $i_D$, $i_q$, $i_Q$, and the control input $v_F$ have been converted to the corresponding root mean square (RMS) quantities $I_d$, $I_F$, $I_D$, $I_q$, $I_Q$, and $V_F$, by substituting $\frac{i_d}{\sqrt{3}}=I_d$, $\frac{i_F}{\sqrt{3}}=I_F$, $\frac{i_D}{\sqrt{3}}=I_D$, $\frac{i_q}{\sqrt{3}}=I_q$, $\frac{i_Q}{\sqrt{3}}=I_Q$, and $\frac{v_F}{\sqrt{3}}=V_F$. Let $x = [I_d, I_F, I_D, I_q, I_Q, \omega, \delta, T_m, G_V]^\mathrm{T}$ be the vector of state variables, $u = [ V_F, u_T]^\mathrm{T}$ the vector of control inputs, and $y = [ V_t, \omega ]^\mathrm{T}$ the vector of outputs, then \eqref{ninthorder} can be written in the usual state-space form as follows:
\begin{equation}
	\scalebox{0.86}
	{$\begin{aligned}
       &\dot{x} = F(x) + G(x)u, \ \ y = H(x),\\
		&F(x) = \\ &\begin{bmatrix}
		F_{11}x_1 + F_{12}x_2+F_{13}x_3+F_{14}x_4x_6+F_{15}x_5x_6 +F_{16}\sin(x_7-\alpha )\\
		F_{21}x_1+F_{22}x_2+F_{23}x_3 + F_{24}x_4x_6 +F_{25}x_5x_6 + F_{26}\sin(x_7-\alpha )\\
		F_{31}x_1+F_{32}x_2+F_{33}x_3+ F_{34}x_4x_6+F_{35}x_5x_6 + F_{36}\sin(x_7-\alpha )\\
		F_{41}x_1x_6+ F_{42}x_2x_6+F_{43}x_3x_6+F_{44}x_4 +F_{45}x_5 + F_{46}\cos(x_7-\alpha )\\
		F_{51}x_1x_6 +F_{52}x_2x_6+F_{53}x_3x_6 +F_{54}x_4+F_{55}x_5 +F_{56}\cos(x_7-\alpha ) \\
		F_{61}x_1x_4 + F_{62}x_2x_4 + F_{63}x_3x_4 +F_{64}x_1x_5+ F_{65}x_6+F_{66}x_8 \\
		x_6 - 1\\
		F_{81}x_8+F_{82}x_9\\
		F_{91}x_6+F_{92}x_9\\
		\end{bmatrix} \\
		&G(x) = \begin{bmatrix} G_{11} & 0 \\ G_{21} & 0 \\ G_{31} & 0 \\0 & 0 \\ 0 & 0\\ 0 & 0\\ 0 & G_{92} \end{bmatrix}   \\
	\end{aligned} $}
	\label{ninthorder1}
\end{equation}
In \eqref{ninthorder1}, $x\in \mathbb{R}^9$, $u\in\mathbb{R}^2$, and $y\in\mathbb{R}^2$. Also, $F(x):D \rightarrow \mathbb{R}^9$, $G(x):D \rightarrow \mathbb{R}^{9\times 2}$, and $H(x):D \rightarrow \mathbb{R}^{9\times 2}$ are sufficiently smooth on a domain $D\subset \mathbb{R}^9$. Thus, the SMIB is a multiple-input and multiple-output (MIMO) system with two inputs: the excitation field voltage $V_F$ and the turbine valve control $u_T$, i.e., $u = [u_1, u_2]^\mathrm{T} = [V_F, u_T]^\mathrm{T}$, and two regulated outputs: the generator terminal voltage $V_t$ and the angular frequency $\omega $, i.e., $y = [y_1, y_2]^\mathrm{T} = [V_t, \omega]^\mathrm{T}$. The generator terminal voltage $V_t$ is computed as $V_t=\sqrt{V^2_d+V^2_q}$, in which the RMS quantities $V_d$ and $V_q$ are 
\begin{equation}
	\begin{aligned}
		V_d= & \ y_{11}I_d+y_{12}I_F+y_{13}I_D+y_{14}I_q\omega +y_{15}I_Q\omega \\
		& +y_{16}\sin(\delta -\alpha )+i_{11}V_F,\\
		V_q= & \ y_{21}I_d\omega +y_{22}I_F\omega +y_{23}I_D\omega +y_{24}I_q+y_{25}I_Q\\
		& +y_{26}\cos(\delta -\alpha ),\\
	\end{aligned}
	\label{truthvt1}
\end{equation} 
where $y_{11}=R_e+L_eF_{11}$, $y_{12}=L_eF_{12}$, $y_{13}=L_eF_{13}$, $y_{14}=L_eF_{14}+L_e$, $y_{15}=L_eF_{15}$, $y_{16}=L_eF_{16}-V_\infty $, $i_{11}=L_eG_{11}$, $y_{21}=L_eF_{41}-L_e$, $y_{22}=L_eF_{42}$, $y_{23}=L_eF_{43}$, $y_{24}=R_e+L_eF_{44}$, $y_{25}=L_eF_{45}$, and $y_{26}=L_eF_{46}+V_\infty $.

The AVR control loop consists of the excitation field voltage $V_F$ as the control input to regulate the generator terminal voltage $V_t$. Furthermore, the turbine valve control $u_t$ and the turbine gate opening $G_V$ work together to control the mechanical power output $P_m$ of the turbine-governor system by modulating the fluid flow or gas intake via the regulated opening of the turbine-governor system's valves and gates. Consequently, the LFC loop consists of the turbine-governor control system to modulate the mechanical power output $P_m$ and, thus, the frequency $\omega $ of the shaft that drives the rotor of the synchronous generator.

\section{Control-Design Model of the SMIB System}

The high-fidelity model of the SMIB system considers the load constraints, the excitation system, the mechanical torque of the turbine-speed governor system, and various effects introduced by the different rotor circuits, i.e., both field and damper-winding effects in the system model. Thus, the complete mathematical description of a large-scale multi-machine power system, which consists of numerous synchronous generators connected to the grid, is exceedingly complex and challenging to model. Simplifications are used to derive a reduced-order CDM of the SMIB system, which is then used to build robust controllers and conduct power grid stability studies. We now present a fifth-order CDM of the SMIB system, which consists of the classical third-order one-axis $E'_q$ model of the synchronous generator \cite{ref27} and a second-order turbine-governor model \cite{ref7}. The classical one-axis model of the generator ignores the damper winding effects, the $d$- and $q$-axis sub-transient flux linkages, and the $d$-axis transient flux linkage effects \cite{ref28}. This reduced-order CDM captures the salient features and characteristics of the complex high-fidelity SMIB system model while neglecting the less significant effects. At the same time, the CDM is simple enough for robust controller design and stability analysis of multi-machine power systems. The system equations for the nonlinear CDM are as follows: 
\begin{equation}
	\begin{aligned}
		&\dot{E}'_q = f_{11}E'_q+f_{12}\cos(\delta -\alpha)
		+f_{13}\sin(\delta -\alpha)+g_{11}E_{FD},\\
		&\dot{\omega } = f_{21}E'^2_q+
		f_{22}E'_q\cos(\delta -\alpha)
		+ f_{23}E'_q\sin(\delta -\alpha)\\
		& \ \ \ \ \ + f_{24}\sin(\delta -\alpha)\cos(\delta -\alpha) +f_{25}\cos^2(\delta -\alpha)\\
		& \ \ \ \ \ + f_{26}\sin^2(\delta -\alpha) +f_{27}\omega + f_{28}T_m,\\      
		&\dot{\delta } = \omega -1,\\
		&\dot{T}_{m} = f_{41}T_{m}+f_{42}G_{V},\\
		&\dot{G}_V = f_{51}\omega +f_{52}G_V+g_{55}u_T,\\
	\end{aligned}
	\label{CDM_one_axis}
\end{equation}
in which  
\begin{equation*}
	\scalebox{0.89}
	{$\begin{aligned}
		&f_{11} = \frac{-(1+\frac{L_2L_1}{M_1})}{\tau '_{d0}}, \ f_{12} = \frac{L_2L_1V_{\infty}}{M_1\tau '_{d0}}, \ f_{13} = \frac{L_2R_1V_{\infty}}{M_1\tau '_{d0}}, \ g_{11} = \frac{1}{\tau '_{d0}},\\
		&L_1 = L_q+L_e, \ L_2 = L_d-L'_d, \ L_3 = L'_d+L_e, \ L_4 = L_q-L'_d,\\
		&M_1 = (r+R_e)^2+(L'_d+L_e)(L_q+L_e), \ R_1 = r+R_e, \ \tau '_{d0}=\frac{L_F}{r_F},\\
		&f_{21} = -\bigg{(}\frac{R_1}{M_1\tau _j}+\frac{L_4L_1R_1}{M^2_1\tau _j}\bigg{)}, \  
		f_{22} = \bigg{(}\frac{R_1}{M_1\tau _j}+\frac{2L_4L_1R_1}{M^2_1\tau _j}\bigg{)}V_\infty,\\
	\end{aligned}$}
	\label{CDM_one_axis_1}
\end{equation*}
\begin{equation*}
	\scalebox{0.89}
	{$\begin{aligned}
		&f_{23} = -\bigg{(}\frac{L_3}{M_1\tau _j}+\frac{L_4L_1L_3}{M^2_1\tau _j}-\frac{L_4R^2_1}{M^2_1\tau _j}\bigg{)}V_\infty, \\ 
		&f_{24} = -\bigg{(}\frac{L_4R^2_1}{M^2_1\tau _j}-\frac{L_4L_1L_3}{M^2_1\tau _j}\bigg{)}V^2_\infty, \ f_{25} = -\bigg{(}\frac{L_4L_1R_1V^2_\infty }{M^2_1\tau _j}\bigg{)},\\
		&f_{26} = \bigg{(}\frac{L_4L_3R_1V^2_\infty }{M^2_1\tau _j}\bigg{)}, \ f_{27} = -\frac{D}{\tau _j}, \ f_{28} = \frac{1}{\tau _j}, \ f_{41} = -\frac{1}{\tau _T},\\
		&f_{42} = \frac{K_T}{\tau _T}, \ f_{51} = -\frac{K_G}{\tau _GR_T}, \ f_{52} = -\frac{1}{\tau _G}, \ g_{55} = \frac{K_G}{\tau _G},\\
		\end{aligned}$}
	\label{CDM_one_axis_2}
\end{equation*}
and $E'_q$ is the $q$-axis voltage behind the transient reactance $L'_d$, where $L'_d=L_d-\frac{(kM_F)^2}{L_F}$. The $q$-axis voltage $E'_q$ cannot be physically measured in a practical system. Let us define the state variables as $x = [E'_q, \omega, \delta, T_m, G_V]^\mathrm{T}$, and the two control inputs as $u = [u_1, u_2]^\mathrm{T} = [E_{FD}, u_T]^\mathrm{T}$. Thus, \eqref{CDM_one_axis} can be written in the usual state-space form as follows:
\begin{equation}
\scalebox{0.9}
{$\begin{aligned}
		&\dot{x} = f(x) + g(x)u, \ \ y = h(x), \ \ \mathrm{in \ which}\\
		&f(x) = \begin{bmatrix} f_{11}x_1+f_{12}\cos(x_3 -\alpha )+f_{13}\sin(x_3 -\alpha )\\
		[f_{21}x^2_1+ f_{22}x_1\cos(x_3 -\alpha )+ f_{23}x_1\sin(x_3 -\alpha )\cdot \cdot \cdot\\
		\cdot \cdot \cdot + f_{24}\sin(x_3 -\alpha )\cos(x_3 -\alpha ) + f_{25}\cos^2(x_3 -\alpha )\cdot \cdot \cdot\\
		\cdot \cdot \cdot + f_{26}\sin^2(x_3 -\alpha )+f_{27}x_2+f_{28}x_4]\\
		x_2-1\\
		f_{41}x_4+f_{42}x_5\\
		f_{51}x_2+f_{52}x_5 \end{bmatrix} \\
		&g(x) =  \begin{bmatrix} g_{11} & 0\\ 0 & 0\\ 0 & 0 \\ 0 & 0\\ 0 & g_{55}\end{bmatrix}.
	\end{aligned}$}
	\label{reduced_order}
\end{equation}
In \eqref{reduced_order}, $x\in \mathbb{R}^5$, $u\in\mathbb{R}^2$, and $y\in\mathbb{R}^2$. Thus, the reduced-order CDM of the SMIB is a MIMO system with two inputs: the excitation field electromagnetic force (EMF) $E_{FD}$ and the turbine valve control $u_T$, and two outputs: the generator terminal voltage $V_t$ and the angular frequency $\omega $. The generator terminal voltage $V_t$ is computed as $V_t=\sqrt{V^2_d+V^2_q}$, in which $V_d$ and $V_q$ are computed as follows:
\begin{equation}
	\begin{aligned}
	    V_d &= V_{d1}E'_q+V_{d2}\cos(\delta -\alpha )+V_{d3}\sin(\delta -\alpha ),\\
		V_q &= V_{q1}E'_q+V_{q2}\cos(\delta -\alpha )+V_{q3}\sin(\delta -\alpha )+E'_q,\\
	\end{aligned}
	\label{reduced_order_2}
\end{equation} 
in which $V_{d1} = -L_q(r+R_e)/((r+R_e)^2+(L'_d+L_e)(L_q+L_e))$,
$V_{d2} = V_\infty L_q(r+R_e)/((r+R_e)^2+(L'_d+L_e)(L_q+L_e))$, 
$V_{d3} = -V_\infty L_q(L'_d+L_e)/((r+R_e)^2+(L'_d+L_e)(L_q+L_e))$,
$V_{q1} = -L'_d(L_q+L_e)/((r+R_e)^2+(L'_d+L_e)(L_q+L_e))$,
$V_{q2} = V_\infty L'_d(L_q+L_e)/((r+R_e)^2+(L'_d+L_e)(L_q+L_e))$, and
$V_{q3} = V_\infty L'_d(r+R_e)/((r+R_e)^2+(L'_d+L_e)(L_q+L_e))$.

\section{Nonlinear Feedback Linearization Control}	
We now propose a nonlinear feedback linearizing controller (NFLC) and an integral-NFLC (INFLC) based on the reduced-order CDM of the SMIB system \eqref{reduced_order}. The NFLC and INFLC controllers use exact feedback linearization \cite{ref25} to transform the nonlinear SMIB power system into two fully linear decoupled subsystems. Linear state feedback controllers are constructed separately for the two decoupled linear subsystems using the linear quadratic regulator (LQR) technique to improve the stability of the closed-loop interconnected SMIB system. The two decoupled linear subsystems obtained using this strategy are independent of the operating points of the power system. As a result, the NFLC and INFLC controllers outperform linear controllers constructed for fully linearized models of the SMIB system operating near the equilibrium point \cite{ref2,ref3,ref4,ref5,ref6}. The procedure for the NFLC/INFLC controller design is discussed in the following steps:
\begin{itemize}
	\item \bf{Step 1: Relative degree computation}
\end{itemize}

To achieve exact feedback linearization for the NFLC and INFLC controllers, the total relative degree ($r$) of the reduced-order CDM of the SMIB system must be identical to the system's order ($n$) \cite{ref25}. By selecting $\tilde{y} = [\tilde{y}_1, \tilde{y}_2]^\mathrm{T} = [\tilde{h}_1(x), \tilde{h}_2(x)]^\mathrm{T} = [\delta, T_m]^\mathrm{T}$ as the system outputs instead of $y = [y_1, y_2]^\mathrm{T} = [V_t, \omega]^\mathrm{T}$, we can achieve exact linearization of the SMIB system with $r = n = 5$. If $[V_t, \omega]^\mathrm{T}$ are chosen as the system outputs, the SMIB system is only partially linearized, and a partial feedback linearization (PFL) controller is used to stabilize the system \cite{ref13,ref14,ref15,ref16,ref17,ref18}. Thus, the PFL controller should stabilize the partially linearized $r$ dynamics and the remaining $n-r$ internal system dynamics. The problem worsens when this strategy is applied to higher-fidelity SMIB system models and interconnected multi-machine power systems with higher degrees of unmodeled internal dynamics. In such a scenario, guaranteeing the stability of the unknown internal dynamics of the interconnected multi-machine power system using the PFL approach is quite challenging. The goal is to linearize the system to the highest possible degree to cancel the nonlinearities and internal dynamics. The relative degree of the system is computed using $[\delta, T_m]^\mathrm{T}$ as the system outputs. For the SMIB power system, the sub-relative degrees $r_1$ and $r_2$ corresponding to the outputs, $\tilde{h}_1(x) = \delta$ and $\tilde{h}_2(x) = T_m$, are calculated as follows:
\begin{equation}
\scalebox{0.9}	
{$\begin{aligned}
		&L_gL^{1-1}_f\tilde{h}_1(x) = L_g\tilde{h}_1(x) = \frac{\partial \tilde{h}_1(x)}{\partial x}g(x) = 0,\\ 
		&L_gL^{2-1}_f\tilde{h}_1(x) = L_gL_f\tilde{h}_1(x) = \frac{\partial (L_f\tilde{h}_1(x))}{\partial x}g(x) = 0,\\ &L_gL^{3-1}_f\tilde{h}_1(x) = L_gL^{2}_f\tilde{h}_1(x) = \frac{\partial (L^2_f\tilde{h}_1(x))}{\partial x}g(x) \neq 0,\\
		&L_gL^{1-1}_f\tilde{h}_2(x) = L_g\tilde{h}_2(x) = \frac{\partial \tilde{h}_2(x)}{\partial x}g(x) = 0,\\ 
        &L_gL^{2-1}_f\tilde{h}_2(x) = L_gL_f\tilde{h}_2(x) = \frac{\partial (L_f\tilde{h}_2(x))}{\partial x}g(x) \neq 0,\\
	\end{aligned}$}
	\label{nfc_0}
\end{equation} 
in which ($L_f\tilde{h}_1(x)$, $L_f\tilde{h}_2(x)$) and ($L_g\tilde{h}_1(x)$, $L_g\tilde{h}_2(x)$) are the Lie derivatives of the functions $\tilde{h}_1(x)$ and $\tilde{h}_2(x)$ along the vector fields $f(x)$ and $g(x)$, respectively. From \eqref{nfc_0}, $L_gL^{r_1-1}_f\tilde{h}_1(x) \neq 0$ and $L_gL^{r_2-1}_f\tilde{h}_2(x) \neq 0$, in which $r_1 = 3$ and $r_2 = 2$. Therefore, the sub-relative degrees corresponding to the outputs $\delta$ and $T_m$ are $r_1 = 3$ and $r_2 = 2$, respectively. The total relative degree $r$ of the reduced-order SMIB system model is computed as follows: 
\begin{equation}
r = \sum^{2}_{i=1} r_i = r_1 + r_2 = 5 = n.
\end{equation}

\begin{itemize}
	\item \bf{Step 2: Nonlinear coordinate transformation and state feedback linearization}
\end{itemize}

For the nonlinear CDM of the SMIB system \eqref{reduced_order}, we assume that the vector fields $f(x):D \rightarrow \mathbb{R}^5$, $g(x):D \rightarrow \mathbb{R}^{5\times 2}$, and the readout map $\tilde{h}(x):D \rightarrow \mathbb{R}^{5\times 2}$ are smooth in the domain $D\subset\mathbb{R}^5$, i.e., their partial derivatives with respect to $x$ of any order exist and are continuous in $D$. Let us define a nonlinear coordinate transformation $z=T(x)$ as
\begin{equation}
	\scalebox{0.96}
	{$\begin{aligned}
		z_1 &= x_3 = \tilde{h}_1(x) = \delta, \\
		z_2 &= \dot{x}_3=\dot{z}_1=\dot{\delta } = L_f\tilde{h}_1(x) =\omega -1,\\
		z_3 &= \ddot{x}_3=\dot{z_2}=\ddot{\delta }= L^2_f\tilde{h}_1(x) = f_{21}E'^2_q+
		f_{22}E'_q\cos(\delta -\alpha)\\
		& \ \ \  + f_{23}E'_q\sin(\delta -\alpha)+ f_{24}\sin(\delta -\alpha)\cos(\delta -\alpha)\\
		& \ \ \ +f_{25}\cos^2(\delta -\alpha) + f_{26}\sin^2(\delta -\alpha)+f_{27}\omega +f_{28}T_m,\\
		z_4 &= x_4 = \tilde{h}_2(x) = T_m,\\
		z_5 &= \dot{x}_4 = \dot{z}_4=  \dot{T}_m= L_f\tilde{h}_2(x) = f_{41}T_{m}+f_{42}G_{V}.\\
	\end{aligned}$}
	\label{nfc1}
\end{equation}
The above state transformations are invertible and exist throughout the generator's normal working region or the domain of stable operation ($0^\circ <\delta <180^\circ $). On differentiating \eqref{nfc1} with respect to time $t$, we get
\begin{equation}
	\begin{aligned}
		\dot{z}_1 &= \dot{x}_3= \dot{\delta }=\omega -1= L_f\tilde{h}_1(x) = z_2,\\
		\dot{z}_2 &= \ddot{x}_3= \ddot{\delta }=\dot{\omega }= L^2_f\tilde{h}_1(x) = z_3,\\ 
		\dot{z}_3 &= \dddot{x}_3=\dddot{\delta }=\ddot{\omega }=\sigma  _1(x)+\gamma  _1(x)E_{FD} = w_1,\\
		\dot{z}_4 &= \dot{x}_4 = \dot{z}_4 = \dot{T}_m= L_f\tilde{h}_2(x) = z_5,\\
		\dot{z}_5 &= \ddot{x}_4 =\ddot{z}_4 =\ddot{T}_m=\sigma  _2(x)+\gamma  _2(x)u_T = w_2.\\
	\end{aligned}
	\label{nfc2}
\end{equation}
In \eqref{nfc1} and \eqref{nfc2}, $x = [E'_q, \omega, \delta, T_m, G_V]^\mathrm{T}$, $\sigma  _1(x) = L^3_f\tilde{h}_1(x)$, $\gamma  _1(x) = L_gL^{2}_f\tilde{h}_1(x)$, $\sigma _2(x) = L^2_f\tilde{h}_2(x)$, and $\gamma  _2(x) = L_gL_f\tilde{h}_2(x)$. Solving for $\sigma  _1(x)$, $\gamma  _1(x)$, $\sigma  _2(x)$, and $\gamma  _2(x)$, after some mathematical computations, we get the following relationships:
\begin{equation}
	\scalebox{0.93}
	{$\begin{aligned}
		&\sigma  _1(x) = p_{31}{x_1}^2+p_{32}x_1\cos(x_3 -\alpha )
		+p_{33}x_1\sin(x_3 -\alpha )
		\\
		& +p_{34}\cos^2(x_3 -\alpha )+p_{35}\sin^2(x_3 -\alpha ) \\
		& +p_{36}\sin(x_3 -\alpha)\cos(x_3 -\alpha )
		+p_{37}x_2 +p_{38}x_4\\
		& +p_{39}x_5 +q_{31}x_1x_2\cos(x_3 -\alpha )+q_{32}x_1x_2\sin(x_3 -\alpha )\\
		& +q_{33}x_2\cos^2(x_3 -\alpha ) +q_{34}x_2 \sin^2(x_3 -\alpha )\\ &+q_{35}x_2 \sin(x_3 -\alpha)\cos(x_3 -\alpha ),\\
	    &\gamma  _1(x) = r_{31}x_1+r_{32}\cos(x_3 -\alpha )+r_{33}\sin(x_3 -\alpha ),\\
         &\sigma  _2(x) = p_{51}x_2 +p_{52}x_4+p_{53}x_5, \ \ \gamma  _2(x) = r_{51},\\
	\end{aligned}$}
	\label{nfc3}
\end{equation}
in which $p_{31}=2f_{11}f_{21}+f_{27}f_{21}$, $p_{32}=2f_{21}f_{12}+f_{22}f_{11}-f_{23}+f_{27}f_{22}$, $p_{33}=2f_{21}f_{13}+f_{22}+f_{23}f_{11}+f_{27}f_{23}$,
$p_{34}=f_{22}f_{12}-f_{24}+f_{27}f_{25}$,
$p_{35}=f_{23}f_{13}+f_{24}+f_{27}f_{26}$,
$p_{36}=f_{22}f_{13}+f_{23}f_{12}+2f_{25}-2f_{26}+f_{27}f_{24}$,
$p_{37}=f^2_{27}$, $p_{38}=f_{27}f_{28}+f_{28}f_{41}$,
$p_{39}=f_{28}f_{42}$, $q_{31}=f_{23}$, $q_{32}=-f_{22}$, 
$q_{33}=f_{24}$, $q_{34}=-f_{24}$, $q_{35}=-2f_{25}+2f_{26}$,
$r_{31}=2f_{21}g_{11}$, $r_{32}=f_{22}g_{11}$, $r_{33}=f_{23}g_{11}$,
$p_{51}=f_{42}f_{51}$, $p_{52}=f^2_{41}$, $p_{53}=f_{41}f_{42}+f_{42}f_{52}$, and $r_{51}=f_{42}g_{55}$.
Applying the state feedback control 
\begin{equation}
	\begin{aligned}
		u_1 &= E_{FD}=\gamma  ^{-1}_1(x)(w_1-\sigma  _1(x)),\\
		u_2 &= u_T=\gamma  ^{-1}_2(x)(w_2- \sigma _2(x)),\\ 
	\end{aligned}
	\label{nfc4a}
\end{equation}
and the state transformation $z=T(x)$ to the nonlinear model of the system \eqref{reduced_order} yields a linear model of the system in the new coordinates as
\begin{equation}
	\dot{z}=Az+Bw,
	\label{nfc4}
\end{equation}
in which $z = [z_1, z_2, z_3, z_4, z_5]^\mathrm{T} \in \mathbb{R}^5$, the linear controller $w = [w_1, w_2]^\mathrm{T} \in\mathbb{R}^2$, and the system matrices $A \in \mathbb{R}^{5\times 5}$, and $B \in \mathbb{R}^{5\times 2}$. Using the coordinate transformation $e = z - z_d$, in which $e = [e_1, e_2, e_3, e_4, e_5]^\mathrm{T} \in \mathbb{R}^5$ are the error variables, and $z_d = [z_{1d}, z_{2d}, z_{3d}, z_{4d}, z_{5d}]^\mathrm{T} \in \mathbb{R}^5$ are the desired values of the state variables $z = [z_1, z_2, z_3, z_4, z_5]^\mathrm{T}$, we transform system \eqref{nfc4} into the error coordinates as follows:
\begin{equation}
     \begin{aligned}
		\dot{e}_1 &= e_2 + z_{2d} - \dot{z}_{1d}, \
		\dot{e}_2 = e_3 + z_{3d} - \dot{z}_{2d},\\
		\dot{e}_3 &= w_1 - \dot{z}_{3d}, \
		\dot{e}_4 = e_5 + z_{5d} - \dot{z}_{4d}, \
		\dot{e}_5 = w_2 - \dot{z}_{5d}.\\
		\end{aligned}
	\label{nfc5}
\end{equation}
The controller is designed such that the state variables $z_1, z_2, z_3, z_4, z_5$ reach their desired steady-state values $z_{1d}, z_{2d}, z_{3d}, z_{4d}, z_{5d}$, i.e., $e = z - z_d = 0$, as $t \to \infty$. Since we want the set point to be an equilibrium, all the derivatives of $z_{1d}$ and $z_{4d}$ are zero. Thus, $\dot{z}_{1d} = z_{2d} = 0, \ddot{z}_{1d} = \dot{z}_{2d} = z_{3d} = 0, \dddot{z}_{1d} = \dot{z}_{3d} = 0, \dot{z}_{4d} = z_{5d} = 0, \ddot{z}_{4d} = \dot{z}_{5d} = 0$. 

\begin{itemize}
	\item \bf{Step 3: Decoupling of the MIMO SMIB system}
\end{itemize}

Next, we decompose system \eqref{nfc5} into two decoupled subsystems, subsystem 1: synchronous generator error subsystem, and subsystem 2: turbine-governor error subsystem, as follows:
\begin{align}
\mathrm{subsystem \ 1:} \ \ \dot{e}_G &= A_G e_G + B_Gw_1, \label{nfc6}\\
\mathrm{subsystem \ 2:} \ \ \dot{e}_T &= A_T e_T + B_Tw_2, \label{nfc7}
\end{align}	
in which $e_G = [e_1, e_2, e_3]^\mathrm{T} \in \mathbb{R}^3$, $e_T = [e_4, e_5]^\mathrm{T} \in \mathbb{R}^2$, and the system matrices $A_G \in \mathbb{R}^{3\times 3}$, $B_G \in \mathbb{R}^{3\times 1}$, $A_T \in \mathbb{R}^{2\times 2}$, and $B_T \in \mathbb{R}^{2\times 1}$ are
\begin{equation}
\begin{aligned}
	A_G=\begin{bmatrix} 0 & 1 & 0\\0 & 0 & 1\\0 & 0 & 0 \end{bmatrix}, \  
	B_G=\begin{bmatrix} 0\\0\\1\end{bmatrix}, \
	A_T=\begin{bmatrix} 0 & 1\\0 & 0\end{bmatrix}, \ 
	B_T=\begin{bmatrix} 0 \\ 1\end{bmatrix} .
\end{aligned}
	\label{nfc8}
\end{equation}
To reduce the steady-state error to zero for subsystem 1, we augment the state equation \eqref{nfc6} with the integral state $e_{iG} = \int e_1\mathrm{d}t = \int (z_1 - z_{1d})dt$, to obtain the
\begin{equation}
	\mathrm{augmented \ subsystem \ 1:} \ \ \dot{\tilde{e}}_G = \tilde{A}_G \tilde{e}_G + \tilde{B}_Gw_1,\\ 
	\label{nfc9}
\end{equation}	
in which $\tilde{e}_G = [e_G, e_{iG}]^\mathrm{T} \in \mathbb{R}^4$, and the augmented system matrices $\tilde{A}_G \in \mathbb{R}^{4\times 4}$, $\tilde{B}_G \in \mathbb{R}^{4\times 1}$ are
\begin{equation}
	\begin{aligned}
		\tilde{A}_G=\begin{bmatrix} A_G & 0\\A_{iG} & 0\end{bmatrix}, \  
		\tilde{B}_G=\begin{bmatrix} B_G\\ 0 \end{bmatrix}, \ A_{iG} = \begin{bmatrix} 1 & 0 & 0 \end{bmatrix}.
	\end{aligned}
	\label{nfc10}
\end{equation}
Similarly, to minimize the steady-state error to zero for subsystem 2, we augment the state equation \eqref{nfc7} with the integral state $e_{iT} = \int e_4\mathrm{d}t = \int (z_4 - z_{4d})dt$, to obtain the
\begin{equation}
	\mathrm{augmented \ subsystem \ 2:} \ \ \dot{\tilde{e}}_T = \tilde{A}_T \tilde{e}_T + \tilde{B}_Tw_2,\\ 
	\label{nfc11}
\end{equation}	
in which $\tilde{e}_T = [e_T, e_{iT}]^\mathrm{T} \in \mathbb{R}^3$, and the augmented system matrices $\tilde{A}_T \in \mathbb{R}^{3\times 3}$, $\tilde{B}_T \in \mathbb{R}^{3\times 1}$ are
\begin{equation}
	\begin{aligned}
		\tilde{A}_T=\begin{bmatrix} A_T & 0\\A_{iT} & 0\end{bmatrix}, \  
		\tilde{B}_T=\begin{bmatrix} B_T\\ 0 \end{bmatrix}, \ A_{iT} = \begin{bmatrix} 1 & 0 \end{bmatrix}.
	\end{aligned}
	\label{nfc12}
\end{equation}

\begin{itemize}
	\item \bf{Step 4: Control law formulation}
\end{itemize}

We next design the inner-loop linear controllers $w_1$ and $w_2$ for the augmented subsystems 1 and 2, respectively. The linear controller $w_1$ for the augmented subsystem 1 \eqref{nfc9} is designed as follows:
\begin{equation}
	w_1 = -\tilde{K}_G \tilde{e}_G = -K_G e_G - K_{iG}e_{iG},\\ 
	\label{nfc13}
\end{equation}	
in which the controller gain matrix $\tilde{K}_G = [K_G, K_{iG}]^\mathrm{T} \in \mathbb{R}^{1\times 4}$, $K_G \in \mathbb{R}^{1\times 3}$, and $K_{iG} \in \mathbb{R}$. Likewise, the linear controller $w_2$ for the augmented subsystem 2 \eqref{nfc11} is designed as follows:
\begin{equation}
	w_2 = -\tilde{K}_T \tilde{e}_T = -K_T e_T - K_{iT}e_{iT},\\ 
	\label{nfc14}
\end{equation}	
in which the controller gain matrix $\tilde{K}_T = [K_T, K_{iT}]^\mathrm{T} \in \mathbb{R}^{1\times 3}$, $K_G \in \mathbb{R}^{1\times 2}$, and $K_{iT} \in \mathbb{R}$.

Note that the time subscript $t$ has been omitted in the equations for brevity. The linear controllers $w_1 = -\tilde{K}_G \tilde{e}_G$ and $w_2 = -\tilde{K}_T \tilde{e}_T$ are designed with optimal control theory using the LQR technique. The LQRs for the two linear controllers are designed to minimize the following quadratic performance measures:
\begin{align}
	J_G &= \int (\tilde{e}^T_G(t)Q_G\tilde{e}_G(t) + w^T_1(t)R_Gw_1(t))dt, \label{nfc15}\\
	J_T &= \int (\tilde{e}^T_T(t)Q_T\tilde{e}_T(t) + w^T_2(t)R_Tw_2(t))dt, \label{nfc16}
\end{align}
in which $Q_G \in \mathbb{R}^{4\times 4}$ and $Q_T \in \mathbb{R}^{3\times 3} \geq 0$ are symmetric positive semi-definite weighting matrices, and $R_G \in \mathbb{R} > 0$ and $R_T \in \mathbb{R} > 0$ are positive definite weights. The weighting matrices are chosen and tuned using a trial-and-error procedure to obtain the desired system response. The controller gain matrices $\tilde{K}_G$ and $\tilde{K}_T$ are
\begin{equation}
	\tilde{K}_G = R^{-1}_G \tilde{B}^T_G P_G, \ \ \tilde{K}_T = R^{-1}_T \tilde{B}^T_T P_T,\\
	\label{nfc17}
\end{equation}	
in which $P_G$ and $P_T$ are the solutions of the algebraic matrix Riccati equations
\begin{align}
	&\tilde{A}^T_G P_G + P_G\tilde{A}_G - P_G\tilde{B}_GR^{-1}_G\tilde{B}_G^TP_G + Q_G = 0, \label{nfc18}\\
	&\tilde{A}^T_T P_T + P_T\tilde{A}_T - P_T\tilde{B}_TR^{-1}_T\tilde{B}_T^TP_T + Q_T = 0. \label{nfc19}
\end{align}
The outer-loop nonlinear feedback controllers, $u_1=E_{FD}$ and $u_2=u_T$, are obtained by substituting \eqref{nfc17}, \eqref{nfc13}, and \eqref{nfc14} into \eqref{nfc4a}. The linear controllers $w_1 = -K_G e_G$ and $w_2 = -K_T e_T$ are used for the NFLC design by utilizing subsystems 1 and 2 without integral augmentation given in \eqref{nfc6} and \eqref{nfc7}, respectively, instead of the augmented subsystems 1 and 2 proposed in \eqref{nfc9} and \eqref{nfc11}, respectively. That is the main difference between the NFLC and INFLC controller designs.

\begin{itemize}
	\item \bf{Step 5: Stability analysis}
\end{itemize}

Substituting the linear controllers $w_1 = -\tilde{K}_G \tilde{e}_G$ and $w_2 = -\tilde{K}_T \tilde{e}_T$ into equations \eqref{nfc9} and \eqref{nfc11}, respectively, we get
\begin{equation}
	\dot{\tilde{e}}_G = (\tilde{A}_G - \tilde{B}_G\tilde{K}_G)\tilde{e}_G, \ \ \dot{\tilde{e}}_T = (\tilde{A}_T - \tilde{B}_T\tilde{K}_T)\tilde{e}_T.
	\label{nfc20}
\end{equation}	
The controller gain matrices $\tilde{K}_G$ and $\tilde{K}_T$ are chosen such that the closed-loop matrices $(\tilde{A}_G - \tilde{B}_G\tilde{K}_G)$ and $(\tilde{A}_T - \tilde{B}_T\tilde{K}_T)$ of the augmented subsystems 1 and 2 are Hurwitz. The closed-loop augmented subsystems 1 and 2 are asymptotically stable with linear controllers $w_1$ and $w_2$, implying that the linear system \eqref{nfc4} is similarly asymptotically stable. Feedback linearization is known to preserve system stability as long as the outer-loop state feedback controllers, $u_1=E_{FD}=\gamma ^{-1}_1(x)(w_1-\sigma _1(x))$, $u_2=u_T=\gamma ^{-1}_2(x)(w_2- \sigma _2(x))$, exactly cancel the system nonlinearities, and the state transformation $z=T(x)$ is invertible in the domain of operation. Since the system is fully linearized using this technique, the internal dynamics of the system are zero. Therefore, we don't need to worry about the stabilization of the internal dynamics of the system, like in the case of the PFL controller \cite{ref13,ref14,ref15,ref16,ref17,ref18}. Thus, we can conclude that the NFLC and INFLC controllers asymptotically stabilize the reduced-order CDM \eqref{reduced_order} of the SMIB system in the original coordinates, and the system states converge to their desired equilibrium values, i.e., $\lim\limits_{t \to \infty} \lVert x(t) - x_d(t) \rVert = 0$.  

\begin{itemize}
	\item \bf{Step 6: The $q$-axis voltage $E'_q$ computation and the correlation between the field voltage $V_F$ and excitation field EMF $E_{FD}$ control inputs}
\end{itemize}

We now provide a mechanism for applying the proposed NFLC and INFLC controllers, designed based on the reduced-order CDM of the SMIB system \eqref{reduced_order}, to the high-fidelity model \eqref{ninthorder1} of the SMIB system. We also provide a technique for reconstructing the non-physical $q$-axis voltage $E'_q$ of the reduced-order CDM from the states of the high-fidelity model. We thus establish the interconnection between the control inputs $V_F$ and $E_{FD}$ for the high-fidelity model and the reduced-order CDM, respectively. Among the unique characteristics that set the suggested controllers apart are the correlation between the control inputs for the two models and the practicality of the control schemes on SMIB models of varying complexities. For the reduced-order CDM \eqref{reduced_order}, the two control inputs are $u = [E_{FD}, u_T]^\mathrm{T}$, whereas for the high-fidelity model \eqref{ninthorder1}, $u = [V_F, u_T]^\mathrm{T}$ is the vector of control inputs. Thus, the proposed NFLC and INFLC controllers cannot be directly applied to the high-fidelity model since the control inputs $V_F$ and $E_{FD}$ are different. The field voltage, $V_F$, is related to the excitation field EMF, $E_{FD}$, by the following expression:
\begin{equation}
	\begin{aligned}
		V_F=&\bigg{(}\frac{r_F}{\omega _RkM_F}\bigg{)}E_{FD} = e_{15}E_{FD}.\\
	\end{aligned}             
	\label{EFD_VF_1}
\end{equation}
In \eqref{EFD_VF_1}, $\omega _R=1$ p.u., and $e_{15}=(\frac{r_F}{\omega _RkM_F})$. Substituting $E_{FD}$ from \eqref{nfc4a} into \eqref{EFD_VF_1}, we obtain
\begin{equation}
	V_F= e_{15}\gamma  ^{-1}_1(x)(w_1-\sigma  _1(x)).
	\label{EFD_VF_2}
\end{equation}
From \eqref{EFD_VF_2}, we can see that the field voltage $V_F$ depends on $\sigma _1(x)$ and $\gamma _1(x)$, which in turn depend on the fictitious state variable $E'_q$, which is neither a physical measurable quantity nor a state variable of the high-fidelity model. Since $E'_q$ cannot be measured using sensors, we cannot apply the field voltage $V_F$ directly to the high-fidelity model without eliminating the state variable $E'_q$ in \eqref{EFD_VF_2}. We solve this problem by evaluating $E'_q$ as a function of the measurable states of the high-fidelity model. The detailed derivation of the reduced-order CDM \eqref{reduced_order} gives $E=E'_q-(L_d-L'_d)I_d$ and $I_d=\frac{-(E'_q-V_{\infty q})(L_q+L_e)-V_{\infty d}(r+R_e)}{(r+R_e)^2+(L'_d+L_e)(L_q+L_e)}$, in which $E$ is the stator air gap RMS voltage in p.u., related to the field current $I_F$ in p.u. Substituting $I_d$ in the expression for $E$, and simplifying the corresponding mathematical equations, we obtain
\begin{equation}
	E=e_{11}E'_q- e_{12}\cos(\delta -\alpha ) -e_{13}\sin(\delta -\alpha ), 
	\label{EFD_VF_3}
\end{equation}
in which 
\begin{equation}
	\scalebox{0.95}{$
		\begin{aligned}	
e_{11} = \bigg{(}1+\frac{L_1L_2}{M_1}\bigg{)}, \
e_{12} = \bigg{(}\frac{L_1L_2V_\infty }{M_1}\bigg{)}, \
e_{13} = \bigg{(}\frac{R_1L_2V_\infty }{M_1}\bigg{)}.		
		\end{aligned}$}\\
	\label{EFD_VF_4}
\end{equation}
The stator air gap RMS voltage, $E$, is related to the field current $I_F$ by the following relationship: $E=\omega _R kM_F I_F = e_{14}I_F$. Substituting $E$ into \eqref{EFD_VF_3}, $E'_q$ can be expressed as a function of the field current $I_F$ and the rotor angle $\delta $, which are state variables of the high-fidelity model, as follows:
\begin{equation}
	E'_q=\frac{e_{14}}{e_{11}}I_F+\frac{e_{12}}{e_{11}}\cos(\delta -\alpha )
	+\frac{e_{13}}{e_{11}}\sin(\delta -\alpha ). 
	\label{EFD_VF_5}
\end{equation}
While the field current $I_F$ can be measured using a sensor, it is not always possible to measure the rotor angle $\delta $ using a sensor. Using the relations for $E$ and $I_F$, after some algebraic manipulations, $E'_q$ can also be written as
\begin{equation}
	E'_q = e_{14}I_F+L_2I_d,
	\label{EFD_VF_6}
\end{equation}
in which $I_F$ and $I_d$ are state variables of the high-fidelity model that can be measured using sensors. The field voltage $V_F$ which is correlated to the excitation field EMF $E_{FD}$ by \eqref{EFD_VF_1}, can be applied to the truth model by using either \eqref{EFD_VF_5} or \eqref{EFD_VF_6} to eliminate $E'_q$ in the expression for $V_F$. Thus, taking into account the high-level systems viewpoint and the circuit theoretical aspects of the SMIB system in the derivation of the different models, we have developed a procedure for designing practical controllers for the SMIB system.

\section{LQG/LTR Control}

The NFLC and INFLC controllers proposed in the previous section require the measurement of rotor angles $\delta$ of synchronous generators, along with other measurable and quantifiable variables, to implement the exact feedback linearization of the system. However, the rotor angles of synchronous generators are not directly available for measurement. Several techniques are available in the literature to measure or estimate the rotor angle \cite{ref13,ref22}, which can be augmented with the proposed NFLC and INFLC controllers. However, practically implementing these techniques on a power grid is not straightforward. Also, it is not always possible to obtain accurate measurements of the rotor angle using these techniques since the sensors introduce noise into the system. Also, the sensors are costly and not easy to operate. Thus, the real-time measurement of rotor angles of synchronous generators is not easily feasible in an interconnected multi-machine power system. To overcome the above problem, we use the speed or angular velocity $\omega$ of the synchronous generator, which can be directly measured using speed sensors, as the system output for the controller design \cite{ref13,ref14,ref15,ref16}. Also, the angular velocity, which is directly related to the derivative of the rotor angle, will provide more damping to the system when used as output feedback. To achieve this goal, we propose an alternative strategy based on the LQG/LTR technique, which employs an enhanced Kalman filter to estimate the rotor angle and the remaining immeasurable states of the SMIB system. The LQG/LTR is a robust linear control approach developed by Doyle \cite{ref26} that uses an appropriately designed Kalman filter to recover the LQR robustness features at the plant input \cite{ref23,ref24}. We now present the LQG/LTR design strategy for the SMIB system.

\begin{itemize}
	\item \bf{Step 1: Linearization of the SMIB system model}
\end{itemize}

The SMIB system's reduced-order CDM \eqref{reduced_order} is linearized around a nominal operating point ($x_0$, $u_0$) via the Taylor series approximation and Jacobian linearization approach. The operating condition, which is a steady-state equilibrium of the system, is attained by the system after all the transients die out or decay to zero. The equilibrium point ($x_0$, $u_0$) is computed by solving the differential equation $\dot{x} = f(x_0) + g(x_0)u_0=0$. The reduced-order linearized model of the SMIB system can be written as follows:
\begin{equation}
	\dot{x}=Ax + Bu + v_1, \ \ y=Cx + Du + v_2,
\label{SMIB_linear}
\end{equation}
in which $x \in \mathbb{R}^5 = [\Delta E'_q, \Delta \omega,  \Delta \delta, \Delta T_m, \Delta G_V]^T$, the control input $u \in \mathbb{R}^2 = [\Delta E_{FD}, \Delta u_T]^\mathrm{T}$, the output $y \in\mathbb{R}^2 = [\Delta V_t, \Delta \omega]^T$, the system matrices $A \in \mathbb{R}^{5\times 5}$, $B \in \mathbb{R}^{5\times 2}$, $C \in \mathbb{R}^{2\times 5}$, and $D \in \mathbb{R}^{2\times 2}$, where $\Delta$ is the deviation from the nominal operating condition, i.e., $\Delta E'_q = E'_q-E'_{q0}$, $\Delta \omega = \omega -\omega _0$, $\Delta \delta = \delta -\delta _0$, $\Delta T_m = T_m-T_{m0}$, $\Delta G_V = G_V-G_{V0}$, $\Delta E_{FD} = E_{FD}-E_{FD0}$, and $\Delta u_T = u_T-u_{T0}$. Moreover, $v_1$ and $v_2$ are fictitious process and sensor noise inputs, respectively. In \eqref{SMIB_linear}, the system matrices ($A$, $B$, $C$, $D$) around the operating point ($x_0$, $u_0$) are
\begin{equation}
	\scalebox{0.95}
	{$\begin{aligned}
A &= \begin{bmatrix} f_{11} & 0 & \frac{\partial f_1}{\partial x_3} & 0 & 0\\ \\
\frac{\partial f_2}{\partial x_1} & f_{27} & \frac{\partial f_2}{\partial x_3} & f_{28} & 0\\ \\
0 & 1 & 0 & 0 & 0\\ \\0 & 0 & 0 & f_{41} & f_{42}\\ \\0 & f_{51} & 0 & 0 & f_{52}\end{bmatrix}, \ \ B = \begin{bmatrix} g_{11} & 0\\ 0 & 0 \\ 0 & 0 \\ 0 & 0 \\ 0 & g_{55} \end{bmatrix},\\
 C &= \begin{bmatrix} T_1 & 0 & T_2 & 0 & 0\\ 0 & 1 & 0 & 0 & 0\end{bmatrix}, \ \ D = \begin{bmatrix} 0 & 0\\ 0 & 0\end{bmatrix},
		\end{aligned}$}
	\label{SMIB_linear1}
\end{equation}
in which $\frac{\partial f_1}{\partial x_3}$, $\frac{\partial f_2}{\partial x_1}$, and $\frac{\partial f_2}{\partial x_3}$ are computed at the operating point $x_0$, where $f_1(x)= f_{11}x_1+f_{12}\cos(x_3 -\alpha )+f_{13}\sin(x_3 -\alpha )$, and $f_2(x) = \ f_{21}x^2_1+ f_{22}x_1\cos(x_3 -\alpha )+ f_{23}x_1\sin(x_3 -\alpha )+ f_{24}\sin(x_3 -\alpha )\cos(x_3 -\alpha ) +f_{25}\cos^2(x_3 -\alpha ) + f_{26}\sin^2(x_3 -\alpha )+f_{27}x_2+f_{28}x_4$. Also, $T_1 = \big{(}\frac{V_{d0}}{V_{t0}}V_{d1}+\frac{V_{q0}}{V_{t0}}V_{q1}+\frac{V_{q0}}{V_{t0}}\big{)}$, $T_2 = \big{(}-\frac{V_{d0}}{V_{t0}}V_{d2}\sin(\delta _\circ -\alpha )+ \frac{V_{d0}}{V_{t0}}V_{d3}\cos(\delta _\circ -\alpha )-
\frac{V_{q0}}{V_{t0}}V_{q2}\sin(\delta _\circ -\alpha )
+\frac{V_{q0}}{V_{t0}}V_{q3}\cos(\delta _\circ -\alpha )\big{)}$. 

\begin{itemize}
	\item \bf{Step 2: Kalman filter and LQG controller design using loop transfer recovery}
\end{itemize}

Through intensive simulations, we discovered the incapability of the Luenberger observer and the LQG controller to reliably estimate the states of the reduced-order CDM of the SMIB system. Also, the Luenberger observer-based LQR and LQG controllers are not robust to variations in the operating conditions, unmodeled internal dynamics, and uncertainties between the reduced-order CDM and the high-fidelity model of the SMIB system. To address this issue and boost the observer's robustness, we developed a robust LQG controller in which the Kalman filter gains are tuned using the loop transfer recovery (LTR) procedure. A gain adjustment design procedure in the time domain, analogous to loop shaping in the frequency domain, is used to adjust the gains of the Kalman filter \cite{ref23,ref26}. This gain adjustment procedure asymptotically achieves the same loop transfer function as a full-state feedback LQR controller while simultaneously improving the observer's robustness. The Kalman filter is designed such that the full-state feedback LQR robustness properties are recovered at the plant input. The Kalman filter provides estimates of the rotor angle $\delta$ and the q-axis voltage $E'_q$ of the SMIB system that cannot be easily measured using sensors. The Kalman filter that produces an estimate of the state, $\hat{x}$, is of the form 
\begin{equation}
	\begin{aligned}
	 \dot{\hat{x}} &= A\hat{x}+Bu+H(y-\hat{y}), \ \hat{y} = C\hat{x}, \ H(q)= \varPsi (q)C^TV^{-1}_2,
	\end{aligned}                         
	\label{LTR1}
\end{equation}	
in which $H(q)$ is the Kalman filter gain parameterized as a function of a scalar variable $q$, and $\varPsi (q)$ is the unique symmetric positive semi-definite solution of the following algebraic matrix Riccati equation
\begin{equation}
	A\varPsi (q)+\varPsi (q)A^T+V_1(q)-\varPsi (q)C^TV^{-1}_2C\varPsi (q)=0.
	\label{LTR2}
\end{equation}
The covariance matrices $V_1(q) \in \mathbb{R}^{5\times 5}$ and $V_2 \in \mathbb{R}^{2\times 2}$, representing the process and measurement noise intensities associated with the signals $v_1$ and $v_2$, respectively, are regarded as design parameters of the Kalman filter. The fictitious noise signals $v_1$ and $v_2$ are introduced to take into account the uncertainties, unmodeled dynamics, and differences between the reduced-order CDM and the actual plant of the SMIB system. We select $V_1(q)=V^T_1(q)>0$ and $V_2=V^T_2>0$ with $(A,V^{\frac{1}{2}}_1(q))$ being stabilizable and $(C,A)$ being observable. The design parameter $V_1(q)$ is chosen as follows:
\begin{equation}
	\begin{aligned}
		V_1(q)=V_{10}+q^2BVB^T, \\
	\end{aligned}              
	\label{LTR3}
\end{equation} 
in which $V_{10} \in \mathbb{R}^{5\times 5}$. The diagonal matrices $V_{10}$ and $V_2$ represent noise intensities appropriate for the nominal high-fidelity SMIB plant model \eqref{ninthorder1}, and $V \in \mathbb{R}^{2\times 2}$ is a positive definite symmetric matrix. The observer gain for $q=0$ corresponds to the nominal Kalman filter gain $H(0)$. By substituting \eqref{LTR3} into \eqref{LTR2} and dividing \eqref{LTR2} by $q^2$, we get
\begin{equation}
	\scalebox{0.95}
	{$\begin{aligned}
	A\frac{\varPsi (q)}{q^2}+\frac{\varPsi (q)}{q^2}A^T+\frac{V_{10}}{q^2}+BVB^T
	-\varPsi (q)C^TV^{-1}_{2}C\frac{\varPsi (q)}{q^2}=0.
    \end{aligned}$}
	\label{LTR4}
\end{equation}
We know that $\frac{\varPsi (q)}{q^2}\rightarrow 0$ as $q\rightarrow \infty$ whenever the transfer function $C(sI-A)^{-1}B$ has no right half-plane zeros. Therefore, as $q\rightarrow \infty$, the gains are seen from \eqref{LTR4} to satisfy 
\begin{equation}
	\scalebox{0.9}
	{$\begin{aligned}
	\varPsi (q)C^TV^{-1}_2C\frac{\varPsi (q)}{q^2}\rightarrow BVB^T, \ \ \mathrm{i.e.,} \ \ \frac{H(q)V_2H^T(q)}{q^2}\rightarrow BVB^T.
    \end{aligned}$}
	\label{LTR5}
\end{equation}
Solutions of \eqref{LTR5} are of the form $\frac{H(q)}{q}\rightarrow BV^{\frac{1}{2}}(V^{\frac{1}{2}}_2)^{-1}$, in which $V^{\frac{1}{2}}$ is the square root of $V$, i.e., $(V^{\frac{1}{2}})^T V^{\frac{1}{2}}=V$, and $V^{\frac{1}{2}}_2$ is the square root of $V_2$. The above solution is a special case of the ideal Kalman filter gain selection $\frac{H(q)}{q}\rightarrow BN$ as $q\rightarrow \infty$ for any non-singular matrix $N$. For the ideal Kalman filter gain scenario, the loop transfer functions for the LQG/LTR and full-state feedback controllers are identical if the Kalman filter dynamics satisfy $H(q)[I + C(sI-A)^{-1}H(q)]^{-1} = B[C(sI-A)^{-1}B]^{-1}$ for all values of the complex variable $s$ \cite{ref26}.

After designing the Kalman filter with the LTR approach, the LQG controller $u = [\Delta E_{FD}, \Delta u_T]^\mathrm{T}$ is designed next with optimal control theory using the LQR technique. The separation principle allows us to design the LQG controller gain $K$ independent of the Kalman filter formulation $H(q)$. The LQG problem is to obtain an optimal control law $u(t)$ that minimizes the following cost function:
\begin{equation}
	\begin{aligned}
		&J_{LQG} = \lim \limits_{T \to \infty} \frac{1}{T}E\bigg{[}\int^T_0 (x^T(t)Qx(t) + u^T(t)Ru(t))dt\bigg{]}, 
	\end{aligned}
	\label{LTR6}
\end{equation}
in which $Q = Q^T\in \mathbb{R}^{5\times 5} \geq 0$ is a symmetric positive semi-definite weighting matrix, and $R = R^T \in \mathbb{R}^{2\times 2} > 0$ is a symmetric positive definite weighting matrix. The matrices $Q$ and $R$ are diagonal matrices with weights appropriately chosen for all the state variables and control inputs to obtain the desired system response. The LQG control law is $u = -K\hat{x}$, in which $K = R^{-1}B^T P$, where $P$ is the unique symmetric positive semi-definite solution of the algebraic matrix Riccati equation $A^TP + PA - PBR^{-1}B^TP + Q = 0$, with $(A, B)$ being controllable and $(Q, A)$ being observable.

\begin{itemize}
	\item \bf{Step 3: Closed-loop stability}
\end{itemize}

Augmenting the SMIB system model \eqref{SMIB_linear} with the Kalman filter dynamics \eqref{LTR1}, and substituting $u = -K\hat{x}$ in equations \eqref{SMIB_linear} and \eqref{LTR1}, the closed-loop dynamics of the SMIB system can be written as follows: 
\begin{equation}
	\begin{aligned}
		\begin{bmatrix} \dot{x} \\ \dot{\hat{x}} \end{bmatrix} = \begin{bmatrix} A & -BK\\H(q)C & A - BK - H(q)C\end{bmatrix} \begin{bmatrix} x \\ \hat{x} \end{bmatrix} + \begin{bmatrix} v_1 \\ H(q)v_2 \end{bmatrix}		 
	\end{aligned}
	\label{LTR7}
\end{equation}
The LQG/LTR controller with $u = -K\hat{x}$ and the Kalman filter gain $H(q)$ designed using the LTR procedure asymptotically stabilizes the SMIB system model \eqref{SMIB_linear} if the closed-loop system dynamics matrix in \eqref{LTR7} is Hurwitz, i.e., the eigenvalues of the control ($A - BK$) and observer ($A - H(q)C$) matrices are in the left half $s$-plane. Since the pair $(A, B)$ is controllable and the pair $(C, A)$ is observable, stability of the closed-loop dynamics \eqref{LTR7} can be ensured by appropriately designing the control ($A - BK$) and observer ($A - H(q)C$) matrices.

To maintain the system's closed-loop stability at every point of the LTR-based Kalman filter adjustment trajectory $H(q)$ with limiting characteristics corresponding to the ideal Kalman filter gain $\frac{H(q)}{q}\rightarrow BN$, extra care should be taken while formulating a stable ($A - H(q)C$) observer dynamics to recover the required robustness at the plant input. The LQG/LTR is a virtual design procedure that uses fictitious noise inputs to represent uncertainties and unmodeled plant dynamics to recover the full-state robustness of minimum phase systems asymptotically at the plant input. At $q = 0$, the Kalman filter will be optimal for the nominally linearized SMIB system \eqref{SMIB_linear} with the system noise signals $v_1$ and $v_2$ known and modeled precisely. Theoretically, as the value of $q$ increases, the noise rejection ability of the filter decreases, but the closed-loop stability margin of the system increases \cite{ref26}, reaching its maximum at $q=\infty$. Nevertheless, simulations confirm that, for large values of $q$, the LQG/LTR controller application to the high-fidelity SMIB plant model \eqref{ninthorder1} leads to instability, as a practical implementation only achieves partial recovery. A full recovery at high values of $q$ will deteriorate the nominal performance of the system and render the system unstable. Thus, the choice of $q$ is critical, and it cannot be arbitrarily selected to a large value when applying this design adjustment procedure to the actual plant. To achieve a reasonable trade-off between the noise/disturbance rejection, closed-loop stability margin, and nominal performance of the system, the scalar $q$ is appropriately tuned by a trial-and-error process. Furthermore, the LQG/LTR design technique is appropriate for robust SMIB system control since the open-loop SMIB system is square, minimum phase, controllable, and observable.

\begin{itemize}
	\item \bf{Step 4: Frequency response analysis}
\end{itemize}

The LQG/LTR controller design is analyzed through frequency domain analysis using the linearized MIMO SMIB system model \eqref{SMIB_linear}. The transfer function of the system \eqref{SMIB_linear} between the control inputs $[\Delta E_{FD}, \Delta u_T]^\mathrm{T}$ and outputs $[\Delta V_t, \Delta \omega]^T$ is $N_R(s) = {\small \begin{bmatrix} N_{11}(s) & N_{12}(s)\\N_{21}(s) & N_{22}(s)\end{bmatrix}} = C(sI - A)^{-1}B$, in which $N_{11}(s)$, $N_{12}(s)$, $N_{21}(s)$, and $N_{22}(s)$ are the individual $5^{th}$-order transfer functions between the two control inputs and two outputs, respectively. Similarly, the LQG/LTR controller gain for the MIMO system can be written in the transfer function form as follows: $K_C(s) = {\small \begin{bmatrix} K_{11}(s) & K_{12}(s)\\K_{21}(s) & K_{22}(s)\end{bmatrix}} = K(sI - A + BK + H(q)C)^{-1}H(q)$, in which $K_{11}(s)$, $K_{12}(s)$, $K_{21}(s)$, and $K_{22}(s)$ are the $5^{th}$-order LQG/LTR controller transfer functions. The resulting loop transfer function (LTF) of the feedback system is $H_{LTF}(s) = K_C(s)N_R(s) = {\small \begin{bmatrix} H_{11}(s) & H_{12}(s)\\H_{21}(s) & H_{22}(s)\end{bmatrix}}$, in which $H_{11}(s) = K_{11}(s)N_{11}(s) + K_{12}(s)N_{21}(s)$, $H_{12}(s) = K_{11}(s)N_{12}(s) + K_{12}(s)N_{22}(s)$, $H_{21}(s) = K_{21}(s)N_{11}(s) + K_{22}(s)N_{21}(s)$, and $H_{22}(s) = K_{21}(s)N_{12}(s) + K_{22}(s)N_{22}(s)$, are $20^{th}$-order individual LTFs of the MIMO system. Conversely, the LQG/LTR controller design based on the linearized version of the high-fidelity SMIB system model \eqref{ninthorder1} will result in $9^{th}$-order LQG/LTR controller transfer functions and exponentially higher-order LTFs, which will further increase the complexity of the closed-loop system. The problem is even more challenging when multiple LQG/LTR controllers are designed for interconnected multi-machine power systems. Thus, the LQG/LTR design based on the reduced-order linearized CDM \eqref{SMIB_linear} of the SMIB system is justified. For maximum robustness recovery of the system at the plant input, the Kalman filter gain $H(q)$ should be designed such that $H_{LTF}(s)$ satisfies $\lim \limits_{q \to \infty}K_C(s)N_R(s) = K(sI-A)^{-1}B$, which corresponds to the ideal Kalman filter gain $H(q) = qBV^{\frac{1}{2}}(V^{\frac{1}{2}}_2)^{-1}$. The mathematical representations of all the transfer functions described in this section are excluded for brevity. 
 
\begin{table}
	\begin{center}
		\caption{Frequency Domain Analysis for Various $q$ Parameter Values.}
		\label{tab1}
		\begin{tabular}{| c | c | c | c | c |}
			\hline
			LQG/LTR  & GM (dB) & PM (deg) & GM (dB) & PM (deg) \\
			Design &  $H_{11}(s)$ &  $H_{11}(s)$ & $H_{22}(s)$ & $H_{22}(s)$\\
			\hline
            $q = 0$ & $\infty$ & 75.616 & 8.347 & $\infty$\\
			\hline
			$q = 9.0005$ & $\infty$ & 71.793 & 45.195 & $\infty$\\
			\hline
			$q = 100$ & 0.0684 & 77.533 & 0.4622 & 69.475\\
			\hline 
			Ideal $H(q)$ & $\infty$ & 69.501 & 3.561 & 36.046\\
            \hline
		\end{tabular}
	\end{center}
\end{table}
  
\begin{figure}[!t]
	\centering
	\includegraphics[width=\columnwidth]{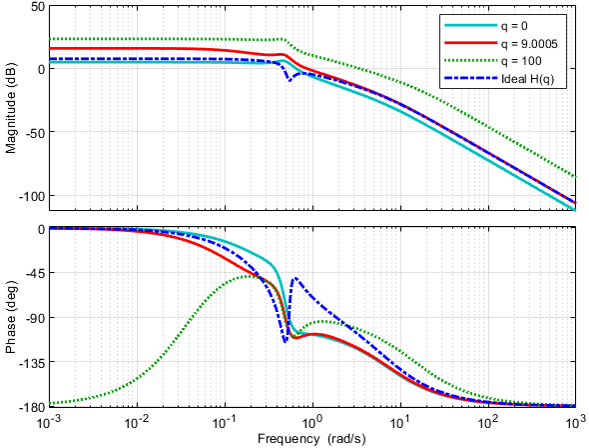}
	\includegraphics[width=\columnwidth]{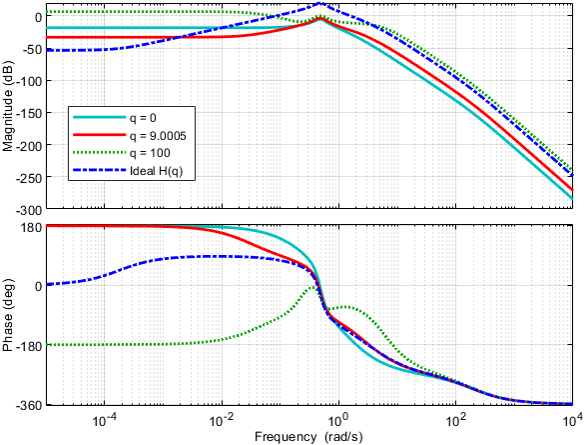}
	\caption{Bode plots (magnitude and phase responses) of the transfer functions $H_{11}$(s) (top) and $H_{22}$(s) (bottom) for various $q$ parameter values.}
	\label{fig3}
\end{figure}

\begin{figure}[!t]
	\centering
	\includegraphics[height = 6cm, width = 8cm]{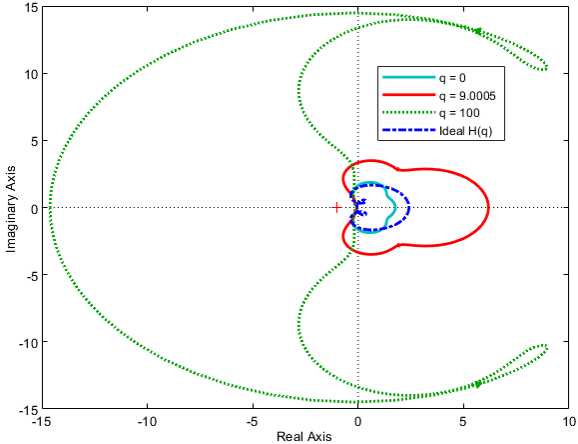}
	\includegraphics[height = 6cm, width = 8cm]{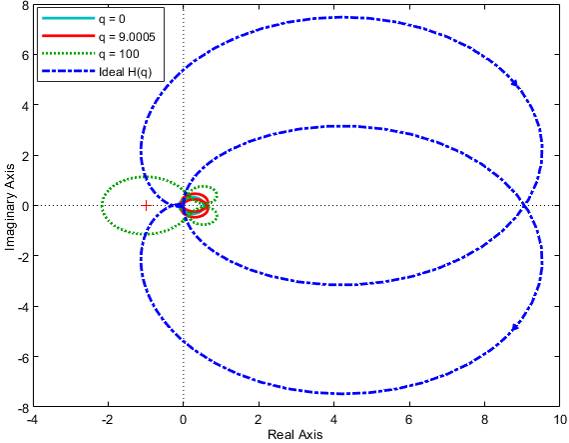}
	\includegraphics[height = 6cm, width = 8cm]{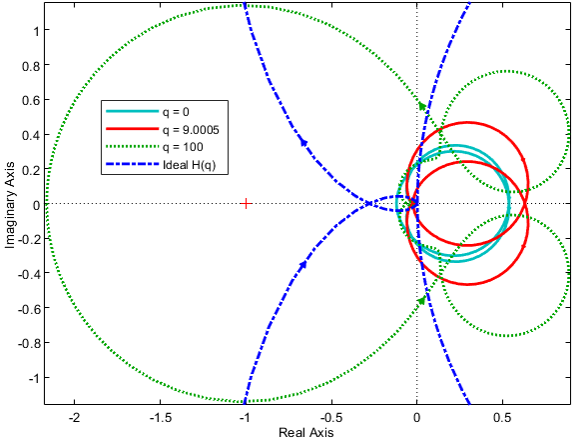}
	\caption{Nyquist diagrams of the transfer functions $H_{11}$(s) (top), $H_{22}$(s) (middle), and $H_{22}$(s) zoomed version (bottom) for various $q$ parameter values.}
	\label{fig4}
\end{figure}

Figs. \ref{fig3} and \ref{fig4} show Bode and Nyquist diagrams of the LTFs $H_{11}$(s) and $H_{22}$(s) for $q$ = 0, 9.0005, 100, and the ideal Kalman filter gain $H(q)$ using the LQR/LTR design procedure. The tuning parameters of the Kalman filter are chosen as $V_{10} = I_{5\times5}$, $V = I_{2\times2}$, and $V_2 = I_{2\times2}$. The LQG controller gain $K$ is designed using the LQR technique \eqref{LTR6} such that the closed-loop system dynamics matrix in \eqref{LTR7} is Hurwitz. Table \ref{tab1} documents the gain margins (GMs) (dB) and phase margins (PMs) (deg) of the LTFs $H_{11}$(s) and $H_{22}$(s) for various $q$ parameter values depicted in the respective Bode and Nyquist diagrams. Figs. \ref{fig3}, \ref{fig4}, and Table \ref{tab1} demonstrate that for the LTF $H_{11}$(s), the GMs are infinite for the ideal $H(q)$, $q$ = 0, and 9.0005; the GM is very low (0.0684 dB) for $q$ = 100; and the PMs range from $69-78$ degrees. Also, for the LTF $H_{22}$(s), the GM of 45.195 dB for $q$ = 9.0005 is significantly higher than the GMs for other $q$ values, and the PMs are infinite for $q$ = 0 and 9.0005. We choose $q$ = 9.0005 for the LQG/LTR controller built for the reduced-order CDM \eqref{reduced_order} to obtain a reasonable compromise between noise/disturbance rejection and robustness recovery within the frequency range of interest. Also, from the Bode plots in Fig. \ref{fig3}, we can see that at higher frequencies, the plots for $q = 9.0005$ and the ideal $H(q)$ converge or move at similar roll-off rates. The Nyquist diagrams of the LTFs $H_{11}$(s) (top), $H_{22}$(s) (middle), and $H_{22}$(s) zoomed version (bottom) in Fig. \ref{fig4} show that the Nyquist contours for $q$ = 0, 9.0005, and the ideal $H(q)$ do not encircle the $-1 + j0$ point. However, the Nyquist contours for $q = 100$ encircle the $-1 + j0$ point once. Since the open-loop SMIB system \eqref{SMIB_linear} is stable, from the Nyquist stability criteria, we can conclude that the closed-loop SMIB system corresponding to the LTFs $H_{11}$(s) and $H_{22}$(s) is unstable for $q = 100$. Thus, we can reaffirm that the closed-loop SMIB system becomes unstable, and the margins approach zero for large values of $q$. The frequency response analyses of the LTFs $H_{12}(s)$ and $H_{21}(s)$ are excluded for brevity.

\section{Simulation and Comparison}

The proposed NFLC, INFLC, and LQG/LTR control strategies are first validated on the reduced-order CDM of the SMIB system \eqref{reduced_order} under different scenarios through the following three case studies: 
\begin{itemize}
	\item Case 1: Nominal operating condition (Operating Point I) for the SMIB CDM
	\item Case 2: A three-phase short-circuit fault at the generator terminal for the SMIB CDM
	\item Case 3: Changes in mechanical power input to the generator for the SMIB CDM
\end{itemize}
Finally, the performance of the controllers is evaluated on the high-fidelity model of the SMIB system \eqref{ninthorder1}, which is treated as the plant model, in the following two operating scenarios:
\begin{itemize}
	\item Case 4: Nominal operating condition (Operating Point I) for the high-fidelity SMIB plant model
	\item Case 5: Increase in the machine loading (Operating Point II) for the high-fidelity SMIB plant model
\end{itemize}

The ratings and parameters of the synchronous generator used in this study are given in Example 4.1 of \cite{ref28}. The parameters of the high-fidelity SMIB power system model in the p.u. system are: $L_d = 1.70$, $L_F = 1.65$, $L_D = 1.605$, $L_q = 1.64$, $L_Q = 1.526$, $kM_F = 1.55$, $kM_D = 1.55$, $M_R = 1.55$, $kM_Q = 1.49$, $r = 0.001096$, $r_F = 0.000742$, $r_D = 0.0131$, $r_Q = 0.0540$, $H = 2.37$ s, $R_e = 0.02$, $L_e = 0.4$, $D = 0$, $\omega_0 = 376.99$ rad, $K_T = 1$, $K_G = 1$, $\tau _T = 0.5$, $\tau _G = 0.2$, $R_T = 20$, $k = \sqrt{3/2}$, $L'_d = 0.245$, $\tau '_{d0} = 5.9$, $\tau _j = 4.74$, $V_\infty = 1$, and $\alpha = 3.5598^\circ $. Similarly, the parameters of the reduced-order SMIB system model in the p.u. system are: $V_{d1} = -0.0249$, $V_{d2} =  0.0249$, $V_{d3} = -0.8037$, $V_{q1} = -0.3797$, $V_{q2} = 0.3797$, $V_{q3} = 0.0037$, $f_{11} = -0.5517$, $f_{12} = 0.3822$, $f_{13} = 0.0037$, $f_{21} = -0.0101$, $f_{22} = 0.0171$, $f_{23} = -0.3269$, $f_{24} = 0.2235$, $f_{25} = -0.0069$, $f_{26} = 0.0022$, $f_{27} = 0$, $f_{28} = 0.2110$, $f_{41} = -2$, $f_{42} = 2$, $f_{51} = -0.2500$, $f_{52} = -5$, $g_{11} = 0.1695$, and $g_{55} = 5$. To ensure realistic control inputs for a practical SMIB power system, the physical limits of the excitation field voltage of the generator are $E_{FD\mathrm{max}} = 5$ p.u. and $E_{FD\mathrm{min}} = -5$ p.u., which match the physical limits of the IEEE Type II and IEEE Type AC4A excitation systems \cite{ref15,ref19}. Also, the physical limits of the turbine valve gate opening are $G_{V\mathrm{max}} = 1.2$ p.u. and $G_{V\mathrm{min}} = 0$ p.u.

\begin{table}
	\begin{center}
		\caption{Operating Points of the SMIB System}
		\label{tab2}
		\begin{tabular}{|c|c|c|}
			\hline
			Variables (p.u.) & Nominal Operating  & Increase in the\\
			 & Condition & Machine Loading\\
			 & (Operating Point I) &  (Operating Point II)\\
			\hline
			$I_{d0}$ & -0.9185 & -1.4281\\
			\hline
			$I_{F0}$ & 1.6315 & 2.37786\\
			\hline
			$I_{D0}$ & $-4.6204\times 10^{-6}$ & 0\\
			\hline   
			$I_{q0}$ & 0.4047 & 0.37472\\
			\hline  
			$I_{Q0}$ & $5.9539\times 10^{-5}$ & 0\\
			\hline 
			$\omega _0$ & 1.0 & 1.0\\
			\hline 
			$\delta _0$ & 1.0 & 0.88676\\
			\hline 
			$T_{m0}$ & 1.0012 & 1.34899\\
			\hline 
			$G_{V0}$ & 1.0012 & 1.34899\\
			\hline
			$V_{q0}$ & 0.9670 & 1.2575\\
			\hline
			$V_{d0}$ & -0.6628 & -0.6130\\ 
			\hline
			$V_{t0}$ & 1.17233 & 1.39899\\ 
			\hline
			Stator current $I_{a0}$ & 1.0037 & 1.4764\\
			\hline
			$V_\infty $ & 1.0 & 1.0\\
			\hline
			$\alpha $ & $3.5598^\circ $ & $3.5598^\circ $\\
			\hline
			$E'_{q0}$ & 1.1925 & 1.6078\\ 
			\hline
			$\tau '_{d0}$ & 5.90 & 5.90\\
			\hline
			Real power P  & 1.0 & 1.3466\\
			\hline
			Power factor $=\cos \phi $ & 0.85 & 0.652\\
			\hline
		\end{tabular}
	\end{center}
\end{table}
The steady-state operating conditions of the SMIB system depend on the parameters of the synchronous generator, turbine-governor system, transmission line, and machine loading. The two different operating conditions at which the controllers are validated on the high-fidelity SMIB plant model for case studies 4 and 5 are given in Table \ref{tab2} \cite{ref28}. At these operating conditions, a detailed comparative performance study is performed on the proposed NFLC, INFLC, LQG/LTR control methods, and the full-state feedback LQR. The proposed controllers are designed based on the reduced-order CDMs \eqref{reduced_order} and \eqref{SMIB_linear} linearized about the nominal operating condition (Operating Point I). At Operating Point I, the machine loading or the real power $P$ generated by the synchronous generator is 1.0 p.u. at 0.85 lagging power factor conditions. The real power $P \ = V_tI_a\cos \phi $ produced by the synchronous generator increases to 1.3466 p.u. as the power factor decreases to 0.652 with an increase in the machine loading at Operating Point II. Also, the stator current $I_a = 1.0037$ p.u. of the synchronous generator at Operating Point I increases to 1.4764 p.u. at Operating Point II when the load on the synchronous generator is increased. Thus, the operating conditions are varied by varying the machine loading, i.e., by increasing or decreasing the load on the generator.
\begin{figure}[!t]
	\centering
	\includegraphics[width=\columnwidth]{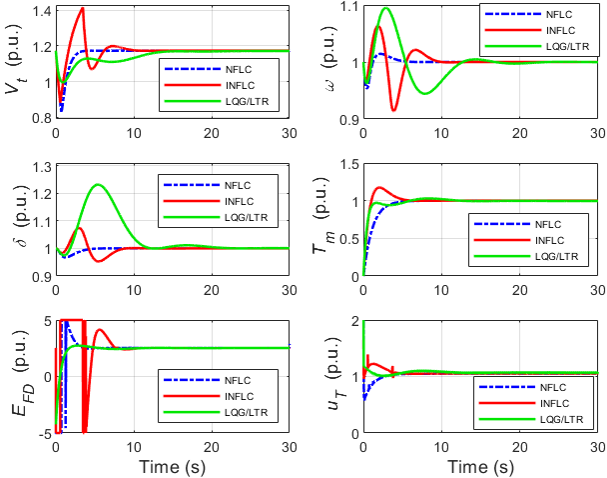}
	\caption{System response plots ($V_t$, $\omega$, $\delta$, $T_m$, $E_{FD}$, $u_T$) for the NFLC, INFLC, and LQG/LTR controllers applied to the reduced-order SMIB CDM at Operating Point I.}
	\label{fig5}
\end{figure}

	\begin{itemize}
		\item \bf{Case 1: Nominal operating condition (Operating Point I) for the SMIB CDM}
	\end{itemize}

The proposed control techniques are first validated on the reduced-order SMIB CDM \eqref{reduced_order} at the nominal operating condition. Fig. \ref{fig5} shows the system response plots for the NFLC, INFLC, and LQG/LTR controllers applied to the reduced-order SMIB CDM at Operating Point I. The simulation results demonstrate that for all the proposed controllers, the outputs $V_t$, $\omega$, $\delta$, and $T_m$ reach their desired steady-state values of 1.17233, 1.0, 1.0, and 1.0012 p.u., respectively. The steady-state equilibrium of the control inputs, $E_{FD(ss)}$ and $u_{T(ss)}$, can be calculated analytically by substituting the steady-state equilibrium values of the state variables $x$ at Operating Point I in the differential equation, $\dot{x}=f(x)+g(x)u=0$, and solving for the control inputs. A closer inspection of Fig. \ref{fig5} indicates that the control inputs, $E_{FD}$ and $u_T$, settle to their desired steady-state values of 2.529 p.u. and 1.0512 p.u., respectively, which conforms with the analytical results. The NFLC approach yields fewer oscillations and smaller peak overshoots of the outputs $V_t$, $\omega$, $\delta$, and $T_m$ compared to the INFLC and LQG/LTR controllers. Also, the NFLC and INFLC controllers hit the maximum physical limit ($\pm 5$ p.u.) of the generator's excitation field EMF $E_{FD}$. In contrast, the LQG/LTR excitation field EMF $E_{FD}$ remains well within the physical limits.
\begin{figure}[!t]
	\centering
	\includegraphics[height=5.4cm,width=\columnwidth]{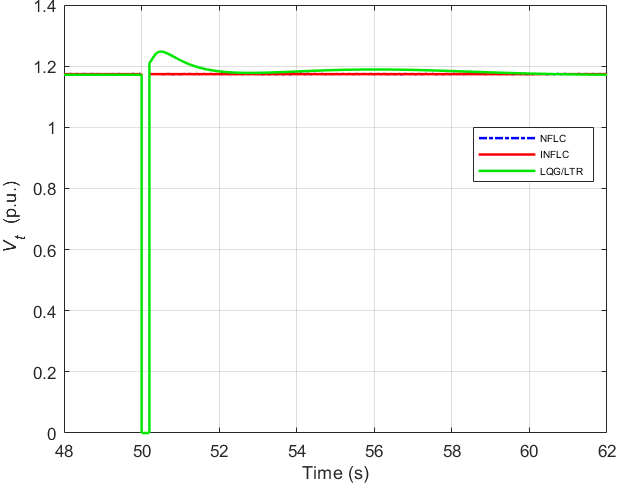}
	\includegraphics[height=5.4cm,width=\columnwidth]{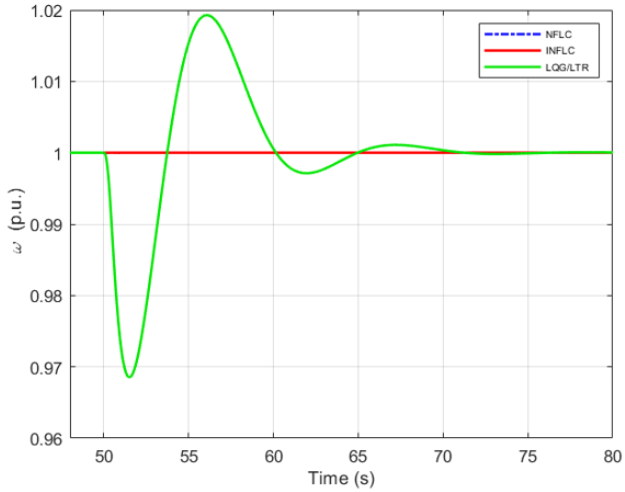}
	\includegraphics[height=5.4cm,width=\columnwidth]{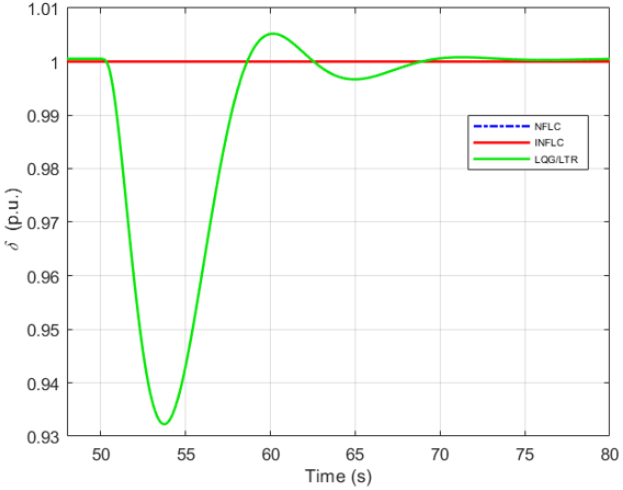}
	\caption{Terminal voltage $V_t$ (top), speed $\omega$ (middle), and rotor angle $\delta$ (bottom) response plots for the NFLC, INFLC, and LQG/LTR controllers applied to the reduced-order SMIB CDM when a three-phase short-circuit fault is applied at the generator terminal.}
	\label{fig6}
\end{figure}
\begin{figure}[!t]
	\centering
	\includegraphics[width=\columnwidth]{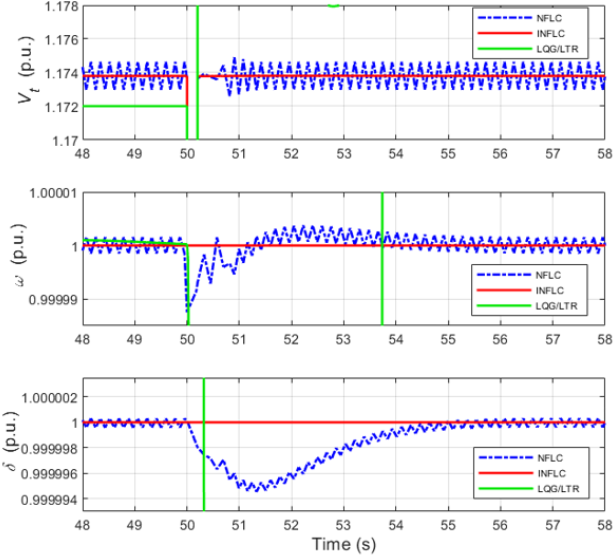}
	\caption{Terminal voltage $V_t$ (top), speed $\omega$ (middle), and rotor angle $\delta$ (bottom) zoomed-in magnified response plots for the NFLC, INFLC, and LQG/LTR controllers applied to the reduced-order SMIB CDM when a three-phase short-circuit fault is applied at the generator terminal.}
	\label{fig7}
\end{figure}
\begin{figure}[!t]
	\centering
	\includegraphics[height=5.4cm,width=\columnwidth]{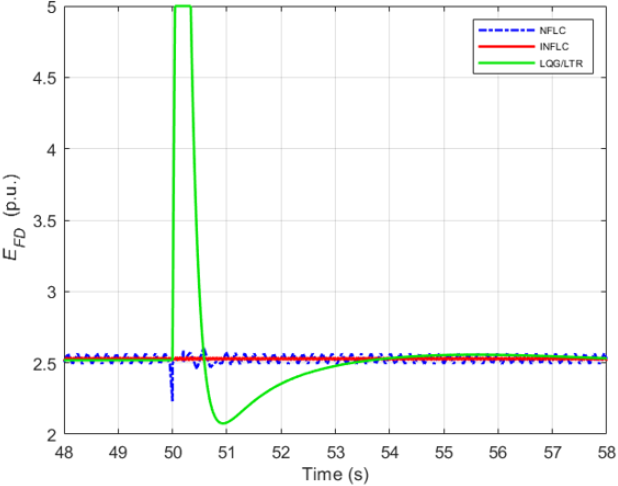}
	\includegraphics[height=5.4cm,width=\columnwidth]{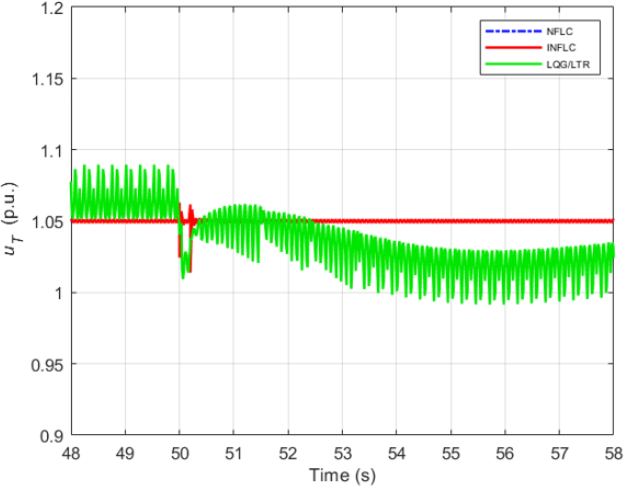}
	\caption{Excitation field EMF $E_{FD}$ (top) and turbine valve control $u_T$ (bottom) magnified response plots for the NFLC, INFLC, and LQG/LTR controllers applied to the reduced-order SMIB CDM when a three-phase short-circuit fault is applied at the generator terminal.}
	\label{fig8}
\end{figure}

	\begin{itemize}
		\item \bf{Case 2: A three-phase short-circuit fault at the generator terminal for the SMIB CDM}
	\end{itemize}

In this case study, the most severe fault in power systems is considered, i.e., a three-phase short-circuit fault is applied at the terminal of the synchronous generator. The following fault sequence is considered to evaluate the performance of the proposed controllers:
\begin{itemize}
	\item The three-phase short-circuit fault occurs at $t = 50$ s.
	\item The fault is cleared at $t = 50.2$ s.
\end{itemize}

Figs. \ref{fig6} to \ref{fig8} show the system response plots for this case study. During the faulted period from $t = 50$ s to $t = 50.2$ s, the generator does not supply any power, and thus, the terminal voltage $V_t$ is zero, as seen in Fig. \ref{fig6} (top). For all the proposed control strategies, the terminal voltage $V_t$ returns to its pre-fault steady-state value after the fault clears at $t = 50.2$ s. However, the LQG/LTR strategy exhibits higher post-fault overshoots in the terminal voltage than the NFLC and INFLC techniques. The synchronous generator operates at a synchronous speed or angular velocity $\omega $ of 1 p.u. for all three controllers during the pre-fault normal operating condition, i.e., until $t = 50$ s, as seen in Fig. \ref{fig6} (middle). However, the speed exhibits more undershoots and overshoots for the LQG/LTR controller than the NFLC and INFLC methods when the fault occurs at $t = 50$ s and after it clears at $t = 50.2$ s. The rotor angle $\delta $ displays similar characteristics for the proposed controllers, with the LQG/LTR controller exhibiting greater undershoots and overshoots than the NFLC and INFLC schemes during the post-fault period, as shown in Fig. \ref{fig6} (bottom). For better resolution of the impact of the fault on $V_t$, it is not on the same time scale as $\omega $ and $\delta $ in Fig. \ref{fig6}.

The LQG/LTR technique can stabilize the SMIB system in the presence of a three-phase short-circuit fault at the generator terminal with some oscillations in the terminal voltage $V_t$, speed $\omega$, and rotor angle $\delta$ responses before they settle to their post-fault steady-state conditions, respectively. However, the nonlinear controllers, NFLC and INFLC, provide more damping into the system compared to the LQG/LTR method and thus respond quicker with faster settling times for the speed $\omega$ and rotor angle $\delta$. Therefore, the nonlinear controllers, NFLC and INFLC, can ensure transient stability of the speed $\omega$ and rotor angle $\delta$ and achieve post-fault steady-state terminal voltage $V_t$ within a few seconds of the faults. The significant difference in performance between the LQG/LTR technique and nonlinear controllers under a three-phase short-circuit fault at the generator terminal can be attributed to the fact that the LQG/LTR technique uses the terminal voltage $V_t$ as the system output for the Kalman filter and controller design. In contrast, the nonlinear controllers utilize the rotor angle $\delta$ as the system output for the controller design, which is less impacted by a three-phase short-circuit fault than the terminal voltage $V_t$.

Fig. \ref{fig7} shows the partially enlarged graphs or zoomed-in magnified versions of the terminal voltage $V_t$, speed $\omega$, and rotor angle $\delta$ responses shown in Fig. \ref{fig6}. The magnified versions of the plots can better distinguish the performance of the two nonlinear controllers, NFLC and INFLC. The terminal voltage $V_t$, speed $\omega$, and rotor angle $\delta$ of the NFLC scheme exhibit more oscillations and take longer times' to settle back to their respective pre-fault steady-state values than the INFLC technique. In the event of a three-phase short-circuit fault at the generator terminal, the INFLC technique eliminates system oscillations and achieves superior transient stability than the NFLC method. The INFLC scheme's integral action rejects system disturbances caused by severe failures while eliminating steady-state errors.

Fig. \ref{fig8} shows magnified plots of the excitation field EMF $E_{FD}$ (top) and turbine valve control $u_T$ (bottom) for the proposed controllers. The LQG/LTR controller hits the maximum physical limit ($\pm 5$ p.u.) of the generator's excitation field EMF $E_{FD}$ before it settles back to its pre-fault steady-state value. The excitation field EMF $E_{FD}$ undergoes small oscillations for the NFLC method, whereas it remains steady for the INFLC scheme. Also, the turbine valve control $u_T$ exhibits higher frequency oscillations for the LQG/LTR technique than the NFLC and INFLC strategies. Slight oscillations in the turbine valve control $u_T$ response are observed for the NFLC and INFLC techniques around the time intervals when the fault is initiated and cleared. The simulation results for Case 2 demonstrate that the INFLC method achieves the best transient stability performance, followed by the NFLC approach, and finally, the LQG/LTR strategy for a severe power system fault. 

\begin{itemize}
	\item \bf{Case 3: Changes in mechanical power input to the generator for the SMIB CDM}
\end{itemize}

The generator's output power varies with the system load, which in turn affects the mechanical power input to the generator through the turbine governor action, adjusting the power output to match the load precisely. The mechanical power input to the generator $P_m$, which is the turbine's output, is numerically equal to the mechanical torque output of the turbine $T_m$ in the p.u. system. In this case study, $P_m$ is held constant at a value of 1 p.u., which is close to the nominal operating condition (Operating Point I), till $t = 25$ s. The mechanical power input to the generator $P_m$ is increased by 10 percent from 1 p.u. to 1.1 p.u. for 25 s, i.e., from $t = 25$ s to $t = 50$ s. Next, $P_m$ is lowered back to 1 p.u. for 25 s, i.e., from $t = 50$ s to $t = 75$ s. Finally, $P_m$ is reduced by 10 percent from 1 p.u. to 0.9 p.u. for 25 s, i.e., from $t = 75$ s to $t = 100$ s. This case study can be summarized as follows in the p.u. system:

\begin{equation}
	\scalebox{0.97}{$
		\begin{aligned}
    P_m = T_m = 
          \begin{cases}
          1, &  0 \ \text{s} \leq t < 25 \ \text{s},\\
          1.1, &  25 \ \text{s} \leq t < 50 \ \text{s},\\
          1, &  50 \ \text{s} \leq t < 75 \ \text{s},\\
          0.9, &  75 \ \text{s} \leq t \leq 100 \ \text{s}.\\
          \end{cases} \\
		\end{aligned}$}
	\label{Pm_var}
\end{equation}

\begin{figure}[!t]
	\centering
	\includegraphics[height=5.4cm,width=\columnwidth]{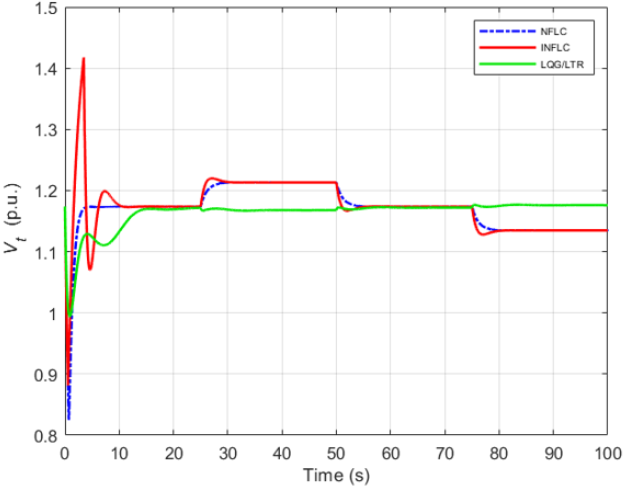}
	\includegraphics[height=5.4cm,width=\columnwidth]{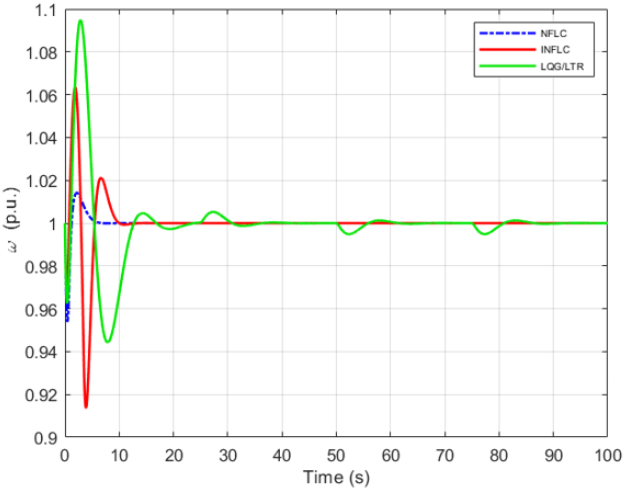}
	\includegraphics[height=5.4cm,width=\columnwidth]{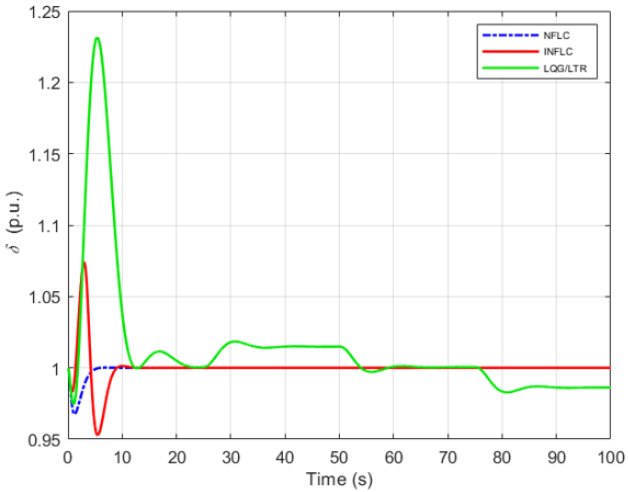}
	\caption{Terminal voltage $V_t$ (top), speed $\omega$ (middle), and rotor angle $\delta$ (bottom) response plots for the NFLC, INFLC, and LQG/LTR controllers applied to the reduced-order SMIB CDM when the mechanical power input to the generator is changed.}
	\label{fig9}
\end{figure}

\begin{figure}[!t]
	\centering
	\includegraphics[height=5.4cm,width=\columnwidth]{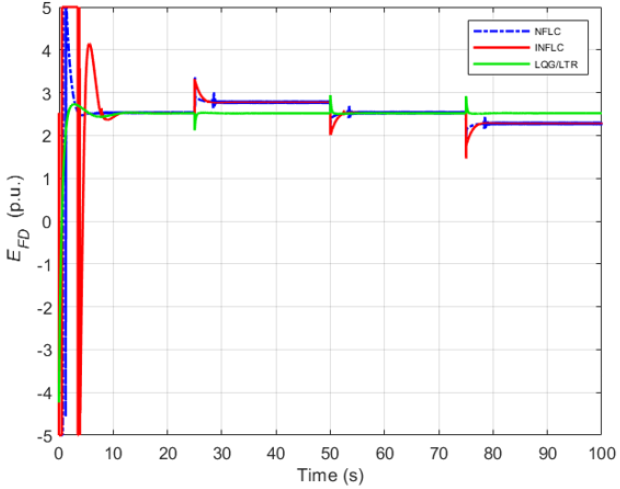}
	\includegraphics[height=5.4cm,width=\columnwidth]{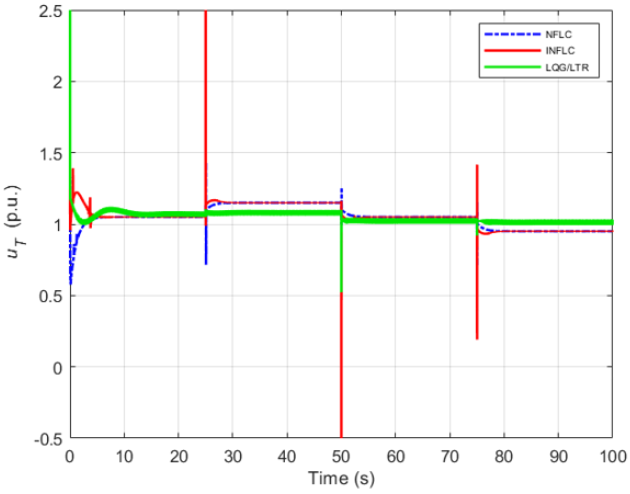}
	\caption{Excitation field EMF $E_{FD}$ (top) and turbine valve control $u_T$ (bottom) response plots for the NFLC, INFLC, and LQG/LTR controllers applied to the reduced-order SMIB CDM when the mechanical power input to the generator is changed.}
	\label{fig10}
\end{figure}

Figs. \ref{fig9} and \ref{fig10} show the system response plots for this case study. Fig. \ref{fig9} (top) shows the terminal voltage $V_t$ response to variations in the mechanical power input $P_m$ for the proposed controllers. The plots show that the NFLC and INFLC controllers undergo step changes in the terminal voltage $V_t$ and settle to new operating points when the mechanical power input $P_m$ is increased ($t = 25$ s to $t = 50$ s) or decreased ($t = 75$ s to $t = 100$ s) from its nominal operating value of 1 p.u. However, the terminal voltage $V_t$ for the LQG/LTR approach remains stable after a brief transient period in response to changes in $P_m$. It settles back near its nominal steady-state operating point. The speed $\omega $ response plots in Fig. \ref{fig9} (middle) for the three controllers reveal that the speed exhibits more undershoots and overshoots before settling to its nominal operating value of 1 p.u. for the LQG/LTR controller than the NFLC and INFLC methods when $P_m$ is changed from its nominal operating value. Fig. \ref{fig9} (bottom) shows that the rotor angle $\delta $ for the LQG/LTR strategy experiences step changes and settles to new operating points when the mechanical power input $P_m$ is varied from its nominal operating value. However, the rotor angle $\delta $ remains stable at its original steady-state equilibrium point for the NFLC and INFLC techniques. 

The distinct system outputs used by the three controllers can explain the step deviations in terminal voltage $V_t$ for the NFLC and INFLC techniques, as well as step changes in rotor angle $\delta $ for the LQG/LTR approach. While the LQG/LTR methodology utilizes $V_t$ and $\omega $ as the outputs and their associated reference values, the NFLC and INFLC techniques use $\delta $ and $T_m$ as the outputs and their corresponding reference values. Furthermore, the generator's output power will increase or decrease as $P_m$ rises or falls for the LQG/LTR method. As the load remains constant, the surplus or deficiency between the power generated and the load will be satisfied by increasing or reducing the speed of the rotating mass of the generator, resulting in the rotor angle $\delta $ stepping up or down and settling to a higher or lower equilibrium point for the LQG/LTR approach. The step changes in terminal voltage $V_t$ to new operating points as $P_m$ increases or decreases for the NFLC and INFLC strategies can be explained similarly.

Fig. \ref{fig10} shows plots of the excitation field EMF $E_{FD}$ (top) and turbine valve control $u_T$ (bottom) for the proposed controllers when the mechanical power input to the generator is changed. These figures illustrate that the INFLC technique exhibits the largest step changes and deviations from the nominal operating condition for the control inputs in response to changes in the mechanical power input. The NFLC approach and the LQG/LTR strategy follow in descending order.

\begin{itemize}
	\bf{\item Case 4: Nominal operating condition (Operating Point I) for the high-fidelity SMIB plant model
	\item Case 5: Increase in the machine loading (Operating Point II) for the high-fidelity SMIB plant model}
\end{itemize}

In case studies 4 and 5, the three proposed controllers' performances are compared to those of a full-state feedback LQR on a high-fidelity SMIB power system plant model at Operating Points I and II. The gains of the proposed controllers tuned for the reduced-ordered CDM are re-tuned for efficient performance on the high-fidelity plant model. Also, the respective controller gains are unchanged for both Operating Points I and II. The LQG/LTR controller gains are chosen as follows: $q$ = 5.25, $V_{10} = I_{5\times5}$, $V = I_{2\times2}$, $V_2 = 0.65I_{2\times2}$, and the controller gain matrix $K=\begin{bmatrix} 87.3944 & -216.7677 & -60.7947 & -13.4353 & -0.0618\\ -1.8244 & 98.0650 & 17.7303 & 42.1399 & 85.8027\end{bmatrix}$. Next, the NFLC controller gains are designed using the LQR technique to obtain $K_G = \begin{bmatrix} 0.09129 & 0.42015 & 0.92121 \end{bmatrix}$ and $K_T = \begin{bmatrix} 0.09129 & 0.43693\end{bmatrix}$. Finally, the INFLC controller gains derived using the LQR technique are as follows: $\tilde{K}_G = \begin{bmatrix} 0.00733 & 0.06795 & 0.36864 & 0.00039 \end{bmatrix}$ and $\tilde{K}_T = \begin{bmatrix} 0.01077 & 0.14674 & 0.00039 \end{bmatrix}$.
\begin{table*}
	\begin{center}
		\caption{Transient and Steady-State Response of Different Controllers at Operating Points I and II}
		\label{tab3}
		\begin{tabular}{|c|c|c|c|c|c|c|}
			\hline
			Control Strategy & Transient Response & Convergence Rate & \multicolumn{2}{c|}{Steady-State Error at Operating Point I} & \multicolumn{2}{c|}{Steady-State Error at Operating Point II}\\
			\hline
			&	 &  & Terminal Voltage $V_t$ & Rotor Angle $\delta $ & Terminal Voltage $V_t$ & Rotor Angle $\delta $ \\
			\hline
			NFLC & Good & Medium & 0.0012 & $\approx$ 0 & 0.00129 & 0.00076 \\
			\hline
			INFLC & Fair & Slow & 0.00233 & $\approx$ 0 & 0.00194 & 0.00684 \\
			\hline   
			LQG/LTR & Excellent & Fast & 0.00013 & 0 & 0.00401 & 0.00854 \\
			\hline
			Full-state LQR & Excellent & Fast & 0.00025 & 0 & 0.00037 & 0.00066 \\
			\hline
		\end{tabular}
	\end{center}
\end{table*}
\begin{figure}[!t]
	\centering
	\includegraphics[width=\columnwidth]{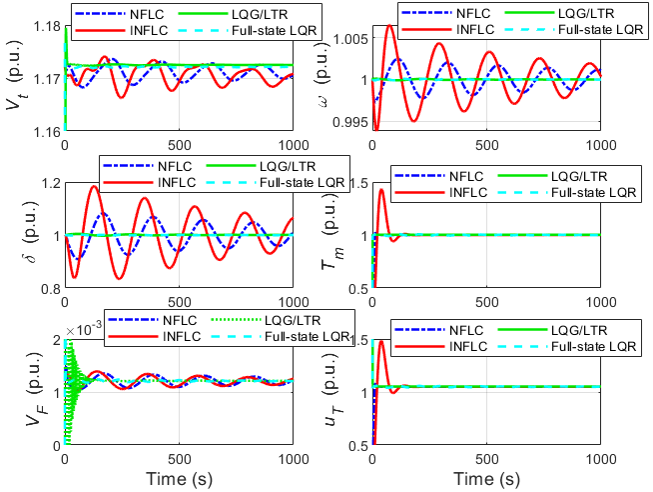}
	\caption{System response plots ($V_t$, $\omega$, $\delta$, $T_m$, $V_F$, $u_T$) for the NFLC, INFLC, LQG/LTR, and full-state LQR controllers applied to the high-fidelity SMIB power system plant model at Operating Point I.}
	\label{fig11}
\end{figure}
\begin{figure}[!t]
	\centering
	\includegraphics[width=\columnwidth]{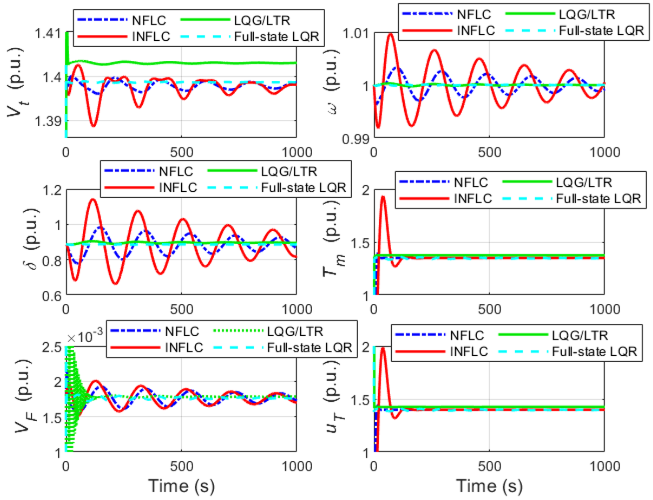}
	\caption{System response plots ($V_t$, $\omega$, $\delta$, $T_m$, $V_F$, $u_T$) for the NFLC, INFLC, LQG/LTR, and full-state LQR controllers applied to the high-fidelity SMIB power system plant model at Operating Point II.}
	\label{fig12}
\end{figure}

Figs. \ref{fig11} and \ref{fig12} show the system response plots for the NFLC, INFLC, LQG/LTR, and full-state feedback LQR controllers applied to the high-fidelity SMIB power system plant model at Operating Points I and II, respectively. Also, Table \ref{tab3} shows the qualitative and quantitative comparison of the transient and steady-state responses of the control strategies at Operating Points I and II. Figs. \ref{fig11} and \ref{fig12} show that $V_t$, $\omega $, $\delta $, and $T_m$ oscillate about their respective steady-state values at the two operating conditions for the four controllers, in which the oscillations decay with time. A closer examination of Fig. \ref{fig11} shows that the generator terminal voltage $V_t$ oscillates about a steady-state value of 1.17113, 1.17, 1.17246, and 1.17208 p.u. for the NFLC, INFLC, LQG/LTR, and full-state LQR controllers, respectively, which deviates from the desired steady-state value of 1.17233 p.u. at Operating Point I. Similarly, a closer view of Fig. \ref{fig12} depicts that the generator terminal voltage $V_t$ oscillates about a steady-state value of 1.3977, 1.39705, 1.403, and 1.39862 p.u. for the NFLC, INFLC, LQG/LTR, and full-state LQR controllers, respectively, which deviates from the desired steady-state value of 1.39899 p.u. at Operating Point II. The angular velocity $\omega $ attains its desired steady-state value of 1.0 p.u. at both operating points, and the rotor angle $\delta $ settles to its desired steady-state value of 1.0 p.u. at Operating Point I for the LQG/LTR and full-state LQR controllers. However, oscillations in $\omega $ and $\delta $ about 1.0 p.u. at Operating Point I are seen for the NFLC and INFLC controllers. Increasing the load on the synchronous generator to 1.3466 p.u. at Operating Point II results in a steady-state error in the rotor angle $\delta $ for all four controllers, as shown in Table \ref{tab3}. A closer look at Fig. \ref{fig12} reveals that the rotor angle $\delta $ oscillates about a steady-state value of 0.886, 0.8936, 0.8953, and 0.8861 p.u. for the NFLC, INFLC, LQG/LTR, and full-state LQR controllers, respectively, which deviates from the desired steady-state value of 0.88676 p.u. at Operating Point II. The generator excitation voltage $V_F$ settles to its steady-state values of 0.00121 p.u. and 0.00176 p.u. at Operating Points I and II, respectively, for the four controllers. The mechanical torque $T_m$ and the turbine valve control $u_T$ settle to their steady-state values of 1.0012 p.u. and 1.0512 p.u., respectively, at Operating Point I for the four controllers, whereas small oscillations about the steady-state values are observed for all four controllers as the machine loading increases at Operating Point II. 

The simulation results for case studies 4 and 5 reveal that the LQG/LTR and full-state LQR controllers exhibit substantially less oscillatory behavior with faster convergence rates than the NFLC and INFLC control approaches, in which the oscillations persist for longer periods with higher amplitudes. Also, the generator terminal voltage $V_t$ exhibits fewer oscillations than the angular velocity $\omega $ and rotor angle $\delta$, as depicted in Figs. \ref{fig11} and \ref{fig12}. The LQG/LTR technique outperforms the NFLC and INFLC strategies in stabilizing the angular velocity $\omega$ and rotor angle $\delta$. The qualitative comparative analysis of the transient responses of the controllers is summarized in Table \ref{tab3}. Also, Table \ref{tab3} shows that the steady-state error of the terminal voltage $V_t$ is lowest for the LQG/LTR controller, followed by the full-state LQR, and highest for the INFLC at Operating Point I. However, the steady-state errors of the terminal voltage $V_t$ and rotor angle $\delta $ are highest for the LQG/LTR controller, followed by the INFLC in descending order, and lowest for the full-state LQR at Operating Point II. The steady-state errors of the NFLC and INFLC controllers are lower than the LQG/LTR control strategy at Operating Point II, as demonstrated in Fig. \ref{fig12}. The NFLC and INFLC controllers perform slightly worse when applied to the high-fidelity plant model since the nonlinear control techniques are model-based and the plant model differs from the reduced-ordered CDM used for the controller design. Therefore, the NFLC and INFLC controllers do not precisely negate the nonlinearities and coupling between the electrical and mechanical dynamics in the SMIB plant model. Overall, the LQG/LTR controller performance is comparable to that of the full-state LQR at various SMIB system operating conditions, with sufficient robustness recovery at the plant input as the LQG/LTR controller's loop transfer function tends toward the full-state LQR loop transfer function. A fair compromise between robustness and closed-loop stability margins can be obtained by appropriately setting the Kalman filter gain $H(q)$, parameter $q$, and covariance matrices of the LQG/LTR controller. 

\section{Conclusion}

In this study, the performance and robustness of the proposed NFLC, INFLC, and LQG/LTR techniques for an SMIB system are evaluated under different operating scenarios. We have presented the physics-based high-fidelity and reduced-order models of the SMIB system and derived the stability and performance criteria for the closed-loop system. Extensive numerical simulations show the advantages and disadvantages of each controller in terms of transient response, steady-state error, disturbance rejection, frequency control, and voltage regulation under various operating scenarios. The simulation results reveal that the INFLC and NFLC approaches outperform the LQG/LTR approach for transient stability under a three-phase short circuit fault at the generator terminal of the SMIB CDM. Furthermore, simulation findings for variations in mechanical power input to the generator for the proposed controllers show that the three controllers perform differently on the SMIB CDM in terms of terminal voltage, speed, and rotor angle responses.

The simulation results for the high-fidelity SMIB plant model under different operating conditions demonstrate that the NFLC and INFLC strategies are more effective at handling the nonlinearities and internal dynamics of SMIB system models than the LQG/LTR approach and nonlinear control techniques like DFL, PFL, backstepping, and adaptive. However, the NFLC and INFLC approaches have certain limitations, including the need for accurate knowledge of system nonlinearities, rotor angle measurement, and sensitivity to noise and disturbances. In contrast, the LQG/LTR controller can provide excellent frequency control, adequate voltage regulation, noise and disturbance rejection, and improved transient stability with guaranteed stability margins regardless of the SMIB system's operating conditions while remaining simple and computationally efficient. The main limitation of the LQG/LTR strategy is its dependence on a stochastic description of the disturbances and a linearized model of the SMIB system around an operating point. The gain scheduling approach can boost the LQG/LTR controller's operating range and robustness by automatically adjusting the controller gains under various SMIB system operating situations. The control synthesis framework and comparative study have highlighted that the choice of the control method depends on the trade-off between performance and complexity, as well as the power system's specific characteristics and restrictions. 

Future work will apply the proposed control techniques to multi-machine power systems, like the IEEE benchmark 10-machine 39-bus New England test system, under severe faults, such as a three-phase short-circuit fault at the generator terminal and the tripping of one of the multi-machine power system's transmission lines.

\end{document}